\documentclass[]{interact}

\usepackage{epstopdf}

\usepackage{subfigure}
\usepackage{array}

\usepackage{float}
\usepackage{orcidlink}
\usepackage{threeparttable}
\usepackage{natbib}
\bibpunct[, ]{(}{)}{;}{a}{}{,}
\renewcommand\bibfont{\fontsize{10}{12}\selectfont}

\usepackage{enumitem} 

\theoremstyle{plain}

\theoremstyle{definition}

\theoremstyle{remark}

\usepackage{xcolor}

\begin{document}

\articletype{}

\title{Urban mobility and learning: analyzing the influence of commuting time on students' GPA at Politecnico di Milano \footnote{\textit{This is an original manuscript of an article published by Taylor \& Francis in Studies in Higher Education on 08 July 2024, available at: \url{https://doi.org/10.1080/03075079.2024.2374005}.}}}

\author{
\name{Arianna Burzacchi\textsuperscript{$*$a} \orcidlink{0000-0001-8284-4909}, Lidia Rossi\textsuperscript{$*$a,b} \orcidlink{0000-0002-6151-2910}, Tommaso Agasisti\textsuperscript{b} \orcidlink{0000-0002-8146-3079}, Anna Maria Paganoni\textsuperscript{a} \orcidlink{0000-0002-8253-3630} and Simone Vantini\textsuperscript{a} \orcidlink{0000-0001-8255-5306} \thanks{$^*$ The authors wish it to be known that, in their opinion, the first two authors should be regarded as joint First Authors. \\A. Burzacchi. Email: arianna.burzacchi@polimi.it; L. Rossi. Email: lidia.rossi@polimi.it}}
\affil{\textsuperscript{a} MOX Laboratory, Department of Mathematics, Politecnico di Milano, Piazza Leonardo da Vinci 32, 20133, Milano, Italy; \\ \textsuperscript{b} Department of Management, Economics and Industrial Engineering, Politecnico di Milano, Via Lambruschini 4/B, 20156, Milano, Italy}
}

\maketitle

\begin{abstract}
Despite its crucial role in students' daily lives, commuting time remains an underexplored dimension in higher education research. To address this gap, this study focuses on challenges that students face in urban environments and investigates the impact of commuting time on the academic performance of first-year bachelor students of Politecnico di Milano, Italy.
This research employs an innovative two-step methodology. In the initial phase, machine learning algorithms trained on privacy-preserving GPS data from anonymous users are used to construct accessibility maps to the university and to obtain an estimate of students' commuting times. In the subsequent phase, authors utilize polynomial linear mixed-effects models and investigate the factors influencing students' academic performance, with a particular emphasis on commuting time. Notably, this investigation incorporates a causal framework, which enables the establishment of causal relationships between commuting time and academic outcomes.
The findings underscore the significant impact of travel time on students' performance and may support policies and implications aiming at improving students' educational experience in metropolitan areas.
The study's innovation lies both in its exploration of a relatively uncharted factor and the novel methodologies applied in both phases.

\end{abstract}


\section{Introduction} 
\label{sec:introduction}

Access to higher education is widely acknowledged as a key determinant of social mobility, economic advancement, and students' personal development. Universities are pivotal institutions that serve as centers of knowledge creation, skill development, and intellectual growth, playing a significant role in shaping societies, producing a qualified workforce, and promoting innovation. Within metropolitan areas, the concentration of universities presents both opportunities and challenges in terms of accessibility. Metropolitan regions are characterized by high population densities, diverse socio-economic profiles, and complex transportation networks. The availability and proximity of universities within these areas shape educational opportunities and may create disparities that impact student outcomes. Understanding the university accessibility patterns and their implications is hence essential for promoting equitable opportunities in students' educational experiences.

Measuring accessibility consists of evaluating the ease with which individuals can reach and utilize university resources. Commuting time, as an indicator of proximity, offers valuable insights into the accessibility of universities within metropolitan areas \citep{geurs2004accessibility, hu2023identifying}. It reflects the travel duration required for students to reach their educational institutions, taking into account factors such as distance, transportation modes, traffic congestion, and infrastructure quality. 
More specifically, commuting time is defined as the amount of time spent on one-way travel from the home location to the study or working institution. 
It has a central role in most lives of students and workers, both due to its daily repetition within the personal routine of individuals, and for the large amount of time taken away from everyday life. \citet{link_eurostat} reports the estimated average commuting time of employed people in 2019 in the EU-27 to be equal to 25 minutes per day. Regarding students, the estimate increases: according to \citet{gwosc2021social}, in the period between 2018 and 2021, the median time of a commute for a European higher education student living in a student university residence was around 15 minutes and reached 40 minutes for students living with their parents.

Among contextual factors that could affect students' school performance \citep{hanushek1979conceptual}, commuting time is a key factor, and its investigation is the major goal of this research. 
The purpose of the study is to shed light on the relationship between commuting time and academic performance of higher education students. 
The case study of the research regards students who are enrolled in engineering bachelor degrees at Politecnico di Milano, a university institution with two campuses located in the city of Milan, Italy.

\begin{figure}[hbt]
    \centering
    \includegraphics[width=\textwidth]{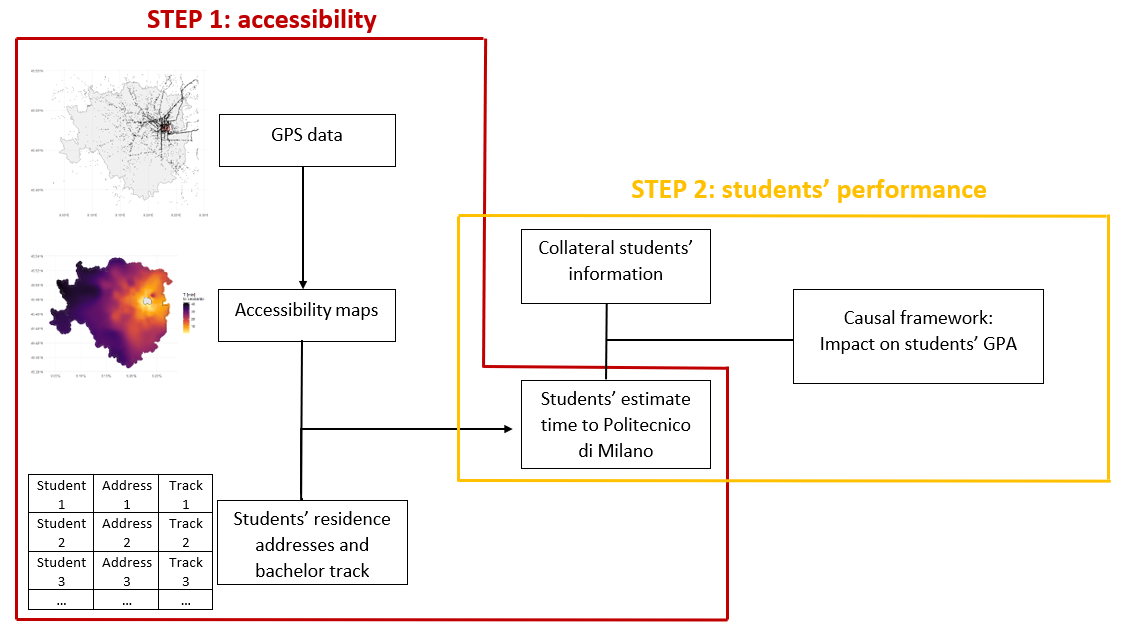}
    \caption{A graphical representation of the analytical framework}
    \label{analytical_framework}
\end{figure}

The research is carried out in a two-step procedure, starting from the estimation of students' commuting time, and moving to the analysis of its impact on students' academic performance. Students' travel time is estimated by means of the non-parametric statistical method of Kernel Regression Estimation using Global Positioning System (GPS) data of anonymous individuals. Here lies one of the main contributions of the research, which combines an innovative data source, i.e. GPS smartphone data, and a data-driven methodology for travel time estimation, overcoming the common limitations found by other studies. In the second phase, the impact of commuting time on academic performance is assessed through the application of balance weighting methods and multilevel polynomial regression models. Students' academic performance is measured through the Grade Point Average (GPA), defined as the average of the marks obtained by the students in the evaluation tests. Other students' information, such as basic generalities (e.g., age, gender) and school-specific features (e.g., bachelor track), are included in the model as confounders to control their effect on academic performances together with commuting time. A multilevel model is used to take into account the variability within different programs and balancing methods are modified accordingly, leading to a methodological innovation in this application field.  According to the authors' knowledge, this is the first study that designs the analysis within a causal framework using balancing methods to evaluate the causal relationship between commuting time and students' performances. A graphical representation of the analytical framework is reported in Figure \ref{analytical_framework}.

The results of this study have the potential to significantly influence the educational experience of students, particularly for those who frequently commute long distances to their campus. Indeed the research highlights the relevance of commuting duration in students' daily encounters and offers insights that could inform policies for the enhancement of students' learning experience.

It is to be noted that the whole analysis is scalable and can be straightforwardly extended to other case studies of university institutions located in other metropolitan cities. 

The paper is organized as follows. A review of the literature is reported in Section \ref{sec:literature}, with both a description of the contextual framework of the research and the analysis of other works with the same goal. Then, methodology and data are explained in Section \ref{sec:method}, while in Section \ref{sec:results} the results are described. Finally, Section \ref{sec:conclusions} is dedicated to conclusions. 

The analysis is carried out by means of the software R \citep{Rsoftware} exploiting some relevant functions from the R packages \texttt{sf, np, lme4, lmerTest, WeightIt}, and \texttt{marginalEffects} \citep{marginaleffects, lme4, WeightIt, np, lmerTest, sf}. The code is available under request to the corresponding authors.

\section{State of the art} 
\label{sec:literature}

\subsection{Conceptual framework - determinants of higher education students performance} \label{subsec:conceptual}

As theorized by \citet{hanushek1979conceptual}, the factors that affect students' school performance are multiple. 
Based on multiple literature reviews \citep{abusaa2019factors, hellas2018predicting, hu2017systematic, kumar2017literature, lei2015academic, rastollo2020analyzing, shahiri2015review}, the determinants of university performance, as assessed by students' evaluations, can be grouped into three categories: personal, socio-economic, and academic factors.

The first set of factors that impact academic success pertains to the demographics of students, such as gender, age, and nationality. 
Such determinants are investigated in observational studies from the literature like \citet{alhajraf2014impact, danilowicz2017determinants, yousef2019determinants}. Personal factors also comprise Motivators, Attitudes, and Beliefs (MAB) factors like fear of punishment and distributive justice, as well as personality traits like the Big Five (Openness, Conscientiousness, Extraversion, Agreeableness, and Neuroticism), as explored in the works of \citet{flom2021helping} and \citet{corazzini2021influence} respectively.

Socio-economic parameters are the second set of factors that impact academic performance. This group pertains to factors related to the family, such as family size, family income, parents’ education and occupation, and residence location. Works investigating the relationship between academic performance and such factors are, for instance, \citet{silva2021influence, rodriguez2020socio, sothan2019determinants}.

Finally, the last group of determinants concerns academic performance parameters. They are classified as institution-related when concerning the place of study both in terms of infrastructures and programs offered, and student-related when regarding students' previous performances and curricula activities. Multiple examples can be found in the literature, for example in the works by \citet{abdelfattah2022predictive, adeyemi2014institutional, alhajraf2014impact, danilowicz2017determinants, gambini2022predictive, HUANG2013133, torenbeek2013predicting, yousef2019determinants, zuluaga2022academic}. The variable under investigation in this study, namely commuting time, falls within the category of institution-related factors, as it serves as a measure of university accessibility. 

\subsection{Evidence of travel time in students' performance} 
\label{subsec:evidence}

Recent studies have highlighted the potential impact of commuting time on students' cognitive functioning, stress, and overall well-being, and its effects are both positive and negative.
On the positive side, commuting time provides individuals time to plan and organize their day, study, or review lessons, and can also serve as a mental break or a relaxing time when to listen to music or read a book. While it can be a time for personal reflection, learning, and development, excessive commuting time has significant drawbacks and negative consequences that cannot be ignored \citep{jachimowicz2021between, pindek2023finally, wilhoit2017my}. First of all, the advantages of travel time are limited in the case of high-quality commute, while are difficult to exploit in most situations, for instance, due to changes in means of transport or to noisy environments. Long and tiring commutes can be physically and mentally exhausting, and limit students in the opportunity to engage in sports or recreational activities, contributing to stress and anxiety \citep{costa1988commuting, hansson2011relationship, kunn2016does}. Given the ambivalent effect of travel time on well-being, it is important to dedicate further attention to in-depth investigation and analysis of this factor. 

For the purposes of this research, an in-depth study is performed of those works that assess the relationship between the accessibility measure of commuting time to educational institutions with students' outcomes. Despite its significance, this topic remains relatively unexplored in the literature, particularly in higher education. It is crucial to delve deeper into this subject to better understand the potential consequences of long commuting times on academic achievement, and ultimately promote students' success in higher education. The examined studies are grouped based on the estimate used for commuting time: geographical distance, estimate from surveys, and distance on the (weighted) road network.

The first measure of commuting time is the geographical distance. This approach assumes that the farther someone lives from their study institution, the longer their daily commute will be. 
The impact of distance between residence and college on educational attainment and GPA is investigated in the work by \citet{garza2018staying}. They focus on a category of disadvantaged students, i.e. first-generation college students, and use a logistic regression model for the binary dependent variable of degree attainment and an Ordinary Least Squares (OLS) regression model for the GPA. Authors know information about ethnicity, age, gender, income, personal and family condition, academic skills, institutional characteristics, school selectivity, and social and academic integration. They find that a greater distance is related to a higher probability of obtaining a bachelor degree but they do not find evidence of a relationship between distance from home and students' GPA. 

Despite its easy computation, a considerable drawback of the accessibility measure through the geographical distance between home and school institution is that it ignores factors such as traffic, transportation options, and the location of essential services, which can have a significant impact on the actual time spent commuting and are not neglectable in the context of metropolitan cities.

Another group of studies analyzes the information collected via questionnaires. This method is used in the works of \citet{kobus2015student} and \citet{tigre2017impact}. 
The first work considers 2,857 students of the Vrije Universiteit in Amsterdam of which they know personal information, commuting time to university, year of attendance, degree program, and context information. The output variable considered is average grades and estimates are computed using OLS and Instrumental Variables (IV). Results of OLS show that the commuting time is not significant, while using IV the commuting time is shown to have a statistically significant negative impact on the outcome.  
The second research uses as output the grade obtained in a standardized mathematics test and collects data of sixth-grade students, their parents and school principals in 118 public schools in Brazil. The authors aim at estimate the impact of commuting time on students' performance using the distance to the closest two schools as IV, and they find strong evidence of a negative effect of time distance on performance. 

Surveys are widely used in all fields of research, and many studies already highlighted the guidelines to avoid response bias. Still, literary works have established that people generally tend to underestimate their travel time, so for that reason, self-reported measurements of commuting time cannot be considered a reliable indicator of actual travel time \citep{stopher2007assessing}.

The last group of works estimates the commuting time as the distance on the road network. Links of the road network are weighted by their expected travel time, usually related to their length, velocity constraints for pavement and street type, or more sophisticated time approximations. The goal of the work by \citet{falch2013geographical} is to investigate the impact of geographical proximity to upper secondary school students on graduation propensity using Norwegian data. They estimate the travel time between students' home ward midpoint and the nearest upper secondary school using ArcGIS Network Analysis. The distance is computed over the public road network and the travel is assumed by car at the speed limits. In the model, they control for personal characteristics, previous grades, socio-economic status, family context, and school environment. Also in this work, the results show a negative impact of travel time on school outcomes. \citet{contreras2018impact} study the causal effect of commuting time over academic achievement on a sample of more than 23,000 eighth-grade students in Santiago. Student's performance is measured as the average score in Mathematics and Spanish on the Sistema de Medición de la Calidad de la Educación (SIMCE) test. Their analysis is performed in a multiple-step process: at the beginning, they compute the travel time from students' home and their university in three modes of transport (by car, by public transport, and by foot) using Google Maps API; then they predict the mode of transport used by the student by means of Machine Learning algorithms, including information about personal characteristics of students and home-school distance; finally, they use OLS, IV and Fixed Effect Models (FEM) to evaluate the effect of the estimated commuting time on the result of SIMCE test.
In the work by \citet{serra2021impact}, employing campus reallocation in Portugal as an exogenous variable, the impact of commuting on students' GPA is estimated by analyzing the changes in commuting times. The administrative dataset encompasses information on 5,546 students who pursued bachelor's degrees at Universidade Nova de Lisboa SBE between the academic years of 2007/08 and 2019/20. The results indicate a negative impact of commuting time on students' GPA. 

These estimates are reliable indicators of commuting time as are able to include traffic conditions and potential delays on the expected travel route. However, they require the specification of the users' mean of transport or the users' average velocity to construct the estimate, so that different values are obtained for different means of transport and speed, and this assumption is a sharp limitation for the assessment of an overall travel time estimate. 

To overcome the exposed limitations in commuting time estimation, the present research proposes the adoption of a new method that makes use of GPS smartphone data of anonymous individuals. GPS traces from smartphones are a relatively recent data source, spreading after the advent of new technologies. They are immensely valuable as they are automatically collected by smartphones effortlessly and at no apparent cost, simultaneously conveying insights about human mobility patterns. Research studies already exploited the valuable information of GPS data to study the human mobility patterns in metropolitan cities \citep{andrade2020mobility, guan2021cuebiq, joseph2020measuring, moro2021mobility}. When assessing the accessibility to healthcare services, mobility data sources such as GPS data have been used for the computation of accessibility maps to hospitals, clinics, and other healthcare facilities, like in \citet{braga2023evaluating} and \citet{xia2019measuring}. To the authors' knowledge, this study is the first work in the education domain that makes use of GPS data for a data-driven commuting time estimation and analysis. 

\section{Methodology and Data} 
\label{sec:method}

\subsection{From GPS data to commuting time estimation}  
\label{subsec:From}

The aim of this step is to build commuting time estimates from any point of origin within the Milan municipality to the university campuses of Politecnico di Milano. These accessibility maps are then used to extract students' estimated commuting time based on their home residence location and their university campus.

Politecnico di Milano has two campuses in Milan: the Leonardo campus, located in a central area of the city, easily reachable by the urban subway and also by suburban and regional trains, and the Bovisa site, which lies in the north-west of the city and is connected by high-frequency suburban trains. Besides the geographical location, the campuses differ by the programs of which they host face-to-face courses, namely each student attends classes in either one of the campuses depending on their program. The commuting time estimation phase took this peculiarity into account, estimating the accessibility maps of both campuses and then extracting each student's commuting time from the map of the sites they visit for classes.

The starting point for the construction of campus accessibility maps is the dataset of smartphone GPS data. The dataset has been provided by Cuebiq Inc., a specialized firm in location intelligence, collecting data in compliance with GDPR privacy regulations, and providing data for research purposes within their ``Data for Good" program. Cuebiq solely collects data from users who consent to anonymized data collection for research purposes through a GDPR compliant opt-in process. The dataset consists of anonymized geo-localization records of mobile phones, containing the location information (latitude and longitude coordinates) at different time instances for Cuebiq users (provided via anonymized IDs). Additional information is also provided, such as the accuracy of the GPS measurement and the type of smartphone operating system that collects the data, as shown in Table \ref{tab::variables_cuebiq}. For the study, coordinates are transformed from the angular longitude and latitude into Universal Transverse Mercator (UTM) coordinates, which use a simpler Cartesian system in meters, to make distance calculations more straightforward compared to the angular measurements of latitude and longitude. In compliance with the research data licensing agreement, at no point did researchers attempt to re-identify anonymized users or trajectories.

\begin{table}[h!]

    \tbl{GPS data description}
    {\begin{tabular}{c p{8cm} c}
    \toprule
    {\textbf{Variable}} &
    {\textbf{Description}} &
    {\textbf{Type}}\\
    \midrule
    Cuebiq ID & Anonymized device ID of Cuebiq user & Cat. \\
    Device type & Mobile operation system (Android, iOS) & Cat. \\
    Timestamp & Date and time in unix format & Num. \\
    Longitude & Coordinate & Num.\\
    Latitude & Coordinate & Num.\\
    Accuracy & Accuracy of the measurement: radius [meters] of the circle around the captured GPS position with a 68\% probability & Num. \\
    \bottomrule
    \end{tabular}}
    \label{tab::variables_cuebiq}
    
\end{table} 

To align with the project objectives, the original dataset is reduced to consider only observations in the period and area of interest. The selected observations are located in the Milan municipality and its surroundings in the period from December 1st to December 20th 2019, i.e. during one month of the first academic semester 2019/20, in working days (from Monday to Friday), and in morning peak hours (arrival between $7$:$30$ and $9$:$30$). The analysis of the reduced dataset allows to capture the human mobility patterns in the city of Milan that mimic the ones of students traveling to reach the university campus for attending the first lesson of the day. The analysis refers to a pre-COVID period, when all the lectures were in presence and students' attendance to face-to-face lectures was wide. The choice of this study period ensures to capture the actual relationship between students' performance and the duration of their home-to-university commute.

With the same goal, another necessary reduction consists in selecting the GPS traces of users whose journey ends at one of the two campuses of Politecnico di Milano, so that the final time estimation would be based on GPS data of actual travelers to Politecnico di Milano. A journey describes the spatial movements of a user between two periods of inactivity through a sequence of GPS data points ordered with respect to the timestamp. The identification of journeys from GPS data passes from the application of some filtering (e.g., remove low-accuracy data) and movement identification criteria (e.g., velocity above a certain threshold), which is described in detail in Appendix \ref{appendix1}. The resulting dataset for what concern the Leonardo campus is composed of 7,670 GPS measurements of 881 trajectories belonging to 448 distinct anonymous Cuebiq users. Regarding the Bovisa campus, there are 4,111 GPS measurements of 498 trajectories belonging to 295 distinct anonymous Cuebiq users. Each of these GPS samples is associated with a measure of commuting time to the university campus, computed as the time elapsed from that measurement to the end of the journey (i.e., the arrival at the university).

To obtain the continuous distribution of commuting time from discrete smartphone GPS data, Kernel Regression Estimation (KRE) is applied. KRE is a non-parametric method to estimate the conditional mean of a response random variable $Y$ conditioning on the values of other random variables $\bm{X}$. In the case of this analysis, the response variable $Y$ represents the commuting time to the university campus and depends on the variables $\bm{X} = [X_1$, $X_2]$ describing the geographical coordinates. The KRE is performed by averaging the values of the response in the observed sample with weights given by kernel function evaluated at the covariate values. Specifically, given a sample of $n$ observations of the response variable $y_i$ and of the covariates $\bm{x}_i = [x_{1i}, x_{2i}]$, the kernel regression estimator of the conditional mean of $Y$ in a new location $\bm{x}$ is expressed as:

\begin{equation}
    \label{eq_1}
    \hat{m} (\bm{x}) = \hat{\mathop{\mathbb{E}}}[Y | \bm{X}=\bm{x}] = \frac{\sum_{i=1}^n{K_H(\bm{x}-\bm{x}_i) \, y_i}}{\sum_{i=1}^n{K_H(\bm{x}-\bm{x}_i)}},
\end{equation}

\noindent
where $K_H$ represents the bidimensional kernel function of bandwidth matrix $H$. The estimator in Equation \ref{eq_1} is the Nadaraya-Watson estimator \citep{nadaraya1964estimating, watson1964smooth}.
Among the various kernel functions the one selected in this work is the Gaussian kernel, as suitable for this applicative problem and as the most commonly used in the literature.

In this work, the symmetry in the two spatial dimensions leads to the definition of the covariance matrix of the kernel as 
a positive scalar $h$ times the $2\!\times\!2$ identity matrix, so that the same amount of smoothing is applied in the two coordinate directions. 
The value $h$ is not held fixed but is varied depending on the location of the estimate and its neighbors, as a generalization of the Nearest-Neighbors (NN) regression. The generalized NN bandwidth is preferred in this work due to the inhomogeneity of the sample data distribution in the space. More formally, $h=h(\bm{x})=c \,d_k(\bm{x})$, where $c$ is a multiplying constant and $d_k$ represents the Euclidean distance between the observation and its $k$-th nearest neighbor. The number of neighbors to be included in the bandwidth $k$ and the multiplying constant $c$ are chosen via the minimization of the Mean Square Error (MSE) of the Leave-One-Out Coss-Validation (LOO-CV) 
on a training set containing data of 85\% trajectories. The remaining trajectory data are used as a test set to evaluate the performance of the method in terms of MSE.

KRE is applied to each of the two campuses of Politecnico di Milano, creating two accessibility maps that indicate the estimated commuting time to university from each point in the Milan municipality. A rectangular bounding box of extension $20.5\!\times\!18.6$km$^2$ is constructed to contain the whole area of the Milan municipality and an additional 1-kilometer wide buffer around it. Inside this box, a grid of 38,316 point locations is defined, with each point positioned 100 meters apart from its adjacent points. The KRE technique is then employed to estimate the commuting time to the university campus for these specified points.

Students' commuting times are determined by geolocating their addresses on the corresponding campus accessibility map and extracting the estimated times accordingly. Specifically, the point on the grid whose distance is the smallest to the residence location of the student is selected. The distance between these points is ensured to be smaller than $50\sqrt{2}$ meters due to the point density of the grid, so the selected point can be considered a good proxy of the student's home location. The estimated commuting time computed for this location is then used as the student's estimated commuting time to university.

\subsection{Investigating the causal relationship between commuting time and academic performance} 
\label{subsec:Investigating}

The analysis focuses on first-year undergraduate students who were enrolled in engineering courses at Politecnico di Milano during the first semester of the academic year 2019/20. The sample is chosen primarily because the study plans exhibit similarity across all programs in this period of interest, then since first-year students are assumed to be more inclined towards attending classes, making them an ideal sample to analyze for the research. 

Another important aspect of the study is its emphasis on students residing in Milan. This selection is motivated by two primary reasons. Firstly, focusing on students within this area enables the construction of accessibility maps with a high degree of precision. Secondly, the study assumes that students residing in Milan experience fewer significant life changes during their first year of university compared to students who relocated far away from their family and friends to pursue higher education, since the latter may face significant challenges that could influence their performance \citep{vieira2018distance}. 
Additionally, the dataset is refined to consider only those students who passed at least one exam at the end of the first semester. This selection is necessary to exclude students for which the GPA cannot be computed due to the absence of exams taken. After the reductions, the dataset comprises 507 observations of first-year students in bachelor engineering programs who both reside in the Milan metropolitan area and passed at least one exam in the analyzed period. The Appendix \ref{appendix2} explores the representativeness of the selected sample. 

The dataset incorporates various personal characteristics of each student, including gender, age of admission to the university, and family income. Additionally, information related to their educational background is considered, such as the final high school grade and the high school track. As for their current academic pursuits, the specific bachelor program followed is taken into account. This last element not only helps identify the campus attended by each student but also serves as a grouping criterion for students. Moreover, the commuting time to university, whose estimate is obtained as in Section \ref{subsec:From}, is considered. In terms of evaluating students' performance, output pertains to the GPA of students at the end of the first semester. The GPA is measured on the Italian grade scale, with marks from 1 to 30 and a threshold of sufficiency of 18. 

To provide a clear overview of the variables involved and their characteristics, as well as to offer descriptive statistics, Table \ref{tab::variables} summarizes all these factors.

\begin{table}[h!]
    \centering
    \footnotesize
    \tbl{Students' variables description and descriptive statistics}{
    \begin{tabular}{p{2.3cm}p{4.7cm}p{2cm}p{3.5cm}}
    \toprule
    \multicolumn{1}{p{2.3cm}}{\textbf{Variable's name}}
    & \multicolumn{1}{p{4.7cm}}{\textbf{Variable's Description}}&\multicolumn{1}{p{2cm}}{\textbf{Typology}}&\multicolumn{1}{p{3.5cm}}{\textbf{Descriptive statistics}}\\
    \midrule  
    Gender & Student's gender& Categorical & Male: 370 (73\%) \textcolor{lightgray}{[base]}
    
    Female: 137 (27\%)\\ &&&\\
    Admission age& Student's age at the admission to university (baseline 18)& Numeric & [-1;14]
    
    mean=0.57
    
    SD=1.11\\ &&&\\
    Family income& Student's family income & Categorical& High: 281 (55\%) \textcolor{lightgray}{[base]}
    
    Middle: 96 (19\%)
    
    Low: 38 (8\%)
    
    Grant: 92 (19\%)\\ &&&\\
    High school grade & Standardized high school final grade & Numeric & [0.6-1]
    
    mean=0.82
    
    SD=0.11\\ &&&\\
    High school track    & High school track attended by the student& Categorical& Humanistic: 29 (6\%)

    Scientific: 444 (88\%) \textcolor{lightgray}{[base]}

    Technical: 26 (5\%)

    Other: 8 (2\%)\\ &&&\\

    Commuting time &Estimated commuting time in the first step expressed in hours & Numeric & [0-0.74]
    
    mean=0.38
    
    SD=0.17\\ &&&\\
    
    GPA & GPA of the student at the end of the first semester of 2019/20 (baseline 18) & Numeric & [0-12]
    
    mean=5.6
    
    SD=3.04  \\
    \bottomrule
    \end{tabular}
    }
    \label{tab::variables}
\end{table} 

The objective of the second step of the analysis is to look for the impact of students' commuting time on their students' performances. The analysis goes beyond the assessment of variable relationships and works in a causal inference domain. The investigation is on the cause-and-effect pattern linking the two variables of interest and representing the variations in the GPA that can be attributed to the variable of commuting time while controlling other factors (e.g., other students' covariates). 

Formalizing the problem, let $Y$ be the outcome variable, here representing the students' GPA. Let then $A$ be the exposure variable, or continuous treatment variable, here describing the students' commuting time, and define the vector of additional covariates of the dataset $\bm{X}$. The interest is in knowing the effect of $A$ on the outcome $Y$ controlling over $\bm{X}$.
This is done by looking at the estimated expected value of the outcome $Y$ for different values $a$ of the exposure variable $A$ , that is, by looking at the estimate of the function $\mu(a) = \mathbb{E}[Y(a)]$. 

The analyzed dataset is observational, meaning it collects measurements of subjects through the external observation of phenomena. Unlike randomized experiments, where treatment can be randomly assigned to subjects, observational studies record data without manipulation and control over them. As a consequence, it is not guaranteed that the relationship that emerges between the outcome and treatment is not affected by other factors that act as confounders. In the case study, for instance, students' characteristics may have an imbalanced distribution between different levels of commuting time, affecting the evaluation of the relationship between the outcome GPA and the commuting time and, hence, working as confounders. Therefore, measuring the impact of the exposure on the outcome is not immediate and comes as a challenge. The Potential Outcome Framework \citep{nayman1938contribution, rubin2005causal} addresses this challenge by formalizing the concept of counterfactuals and offering a systematic way to estimate causal effects in dataset. Within this framework, the Propensity Score and its generalization to the continuous treatment case are the starting point of many methods for observational studies that naturally deal with confounding \citep{hirano2004propensity, rosenbaum1983propensity}. The Generalized Propensity Score summarizes the information about potential confounders into a single score by measuring how likely the combination of exposure and characteristics of a sample unit is represented in the analyzed sample. To formalize its definition, let $r$ denote the conditional density of the continuous treatment variable given the covariates, $r = r(a,\bm{x}) = f_{A|\bm{X}}(a|\bm{x})$. Then, the Generalized Propensity Score is $R = r(A, \bm{X})$.

Among the four classes of Propensity Score techniques used to remove the effects of confounding outlined by \citet{austin2011introduction}, balancing weighting methods are the most popular for dealing with the case of continuous exposure, moving forward from the classic case of binary treatment. These methods work by assigning a weight to each observation in the dataset, measuring the importance that the instance needs to have in the analyzed dataset to make it fully balanced: larger weights are given to subjects who are less common in the study population, ensuring their influence is appropriately accounted for. The state-of-the-art techniques for the computation of balancing weights work by estimating the Generalized Propensity Score and using its inversion as a weight. However, in the last ten years, the goal of covariate balancing has been addressed with other innovative methodologies that either integrate the Generalized Propensity Score estimation in wider covariate balancing methods or assess the dataset balance by looking only at the covariate distribution.

The present research follows the newest works in the field of balancing methods addressing the impact of continuous treatment on the outcome variable. Different weighting methods are explored and compared to find the most suitable for the application. Subsection \ref{subsubsec:weighting} is dedicated to the description of such weight comparison and application. Then, the selected weights are applied to the analyzed sample and the impact of commuting time on students' GPA is investigated, as discussed in Subsection \ref{subsubsec:impact}.

\subsubsection{Dataset balancing}
\label{subsubsec:weighting}
The goal of this subsection is to remove the effect of external confounding factors. Apart from the commuting time and student's GPA, which here work as continuous exposure $A$ and outcome $Y$ respectively, all the observed covariates reported in Table \ref{tab::variables} are included as confounders. It is assumed that no other unobserved determinants of students' performances confound the relationship, since the available covariates already describe students' characteristics from the point of view of the three classes of determinants explained in Section \ref{sec:literature}. 

For the selection of the proper methodology, four weighting methods already available in the literature are compared with respect to their balancing performance in this application case. Specifically, this research employs and compares the Inverse Probability Weights computed from the Generalized Propensity Score, 
the application of Covariate Balancing Propensity Score, both in the parametric and non-parametric case, and the weight estimation from the Entropy Balancing method.

In the first approach, following the path of \citet{rosenbaum1983propensity} and the subsequent generalization of \citet{hirano2004propensity} which deals with the case of continuous exposure, Inverse Probability Weights (IPWs) are computed from the inversion of Generalized Propensity Scores of the observed data. 
The IPWs are obtained in such a way that higher weights are associated with less-represented sample units and vice-versa, balancing the distribution of confounding characteristics across exposure levels. 

The second weighting method analyzed is the Covariate Balancing Propensity Score (CBPS). Proposed by \citet{imai2014covariate}, the method parallelly estimates the balancing weights and models the relationship between exposure and confounders. It is able to overcome the limitation of the classic IPWs of suffering for substantial bias in the estimation after slight misspecification of the Generalized Propensity Score model. Both the parametric and non-parametric versions of the CBPS are used and compared for weight computation \citep{fong2018covariate}.

Finally, the Entropy Balancing (EB) weighting method is considered \citep{hainmueller2012entropy, vegetabile2021nonparametric}. Differently from previously mentioned methods, EB weights are computed without a direct estimation of the generalized propensity score, but through the minimization of a measure of the dispersion of the weights, that is the entropy, while controlling conditions on specified moments of the covariates. The method removes the requirement of modeling the generalized propensity score and shows performance comparable to the one of CBPS.

Most research works on datasets of independent observations with a non-hierarchical structure, for which the theoretical and practical framework of generalized propensity score and balancing methods is already developed and consolidated.
However, in this research application, data have a hierarchical structure (i.e., students belong to different bachelor programs). As pointed out by many as \citet{arpino2011specification} and \citet{li2013propensity}, the clustered nature of the data is a source of relevant information and should not be discarded but integrated into the weighting method. For the case of continuous exposure, \citet{schuler2016propensity} propose to include the grouping factor in the model for generalized propensity score estimation. They use both linear Fixed Effects Models (FEMs) and Random Effects Models (REMs) for modeling the exposure variable from the confounders and getting an estimate of the generalized propensity score and of the IPWs. 

Starting from these studies, the current analysis includes the bachelor program information so as not to neglect the hierarchical structure of the data. Previously mentioned methods are modified: for the computation of IPWs, the program is added either as a fixed effect or as a random effect in the formulation of the generalized propensity score (i.e., through FEMs and REMs); for the CBPS methods, the program is added as a fixed effect in the formulation of the generalized propensity score as in FEMs; likewise, for the EB method, dummy covariates of the bachelor program are added as confounders, in accordance with the FEM approach.

More specifically for what concerns the IPW computation, the model formulation for generalized propensity score estimation is expressed in Model \ref{eq:FEM} for FEM and Model \ref{eq:REM} for REM. 

\begin{equation}\label{eq:FEM} 
    t_{il} = \gamma_0 + \sum_{j=1}^{J-1} \gamma_j x_{jil} + \sum_{l=1}^L z_l + \epsilon_{il}, \:\:\:\: \epsilon_{il} \sim \mathcal{N} (0, \sigma_{\epsilon_1}^2)
\end{equation}
\begin{equation}\label{eq:REM} 
    t_{il} = \gamma_0 + \sum_{j=1}^{J-1} \gamma_j x_{jil} + z_l + \epsilon_{il}, \:\:\:\: z_l \sim \mathcal{N}(0, \sigma_{z}^2), \:\:\:\: \epsilon_{il} \sim \mathcal{N} (0, \sigma_{\epsilon_2}^2)
\end{equation}

In the formulas \ref{eq:FEM} and \ref{eq:REM}, $t_{il}$ indicates the commuting time of student $i \in \{1,\dots,n_l\}$, $n = \sum_l n_l$, in program $l \in \{1,\dots,L\}$; ${\bm{\gamma}=\{\gamma_0,\dots,\gamma_{J-1}\}}$ is the ${J-}$dimensional vector of parameters; $x_{jil}$ represents the value of the $j-$th confounding student covariate;  $z_{l}$ is the either fixed (in FEM) or random (in REM) intercept of the program $l$; $\epsilon_{il}$ is the error; in REM it is assumed that $\bm{z}$ is independent of $\bm{\epsilon}$. 

After their computation, weights are compared following the guidelines proposed in the literature \citep{vegetabile2021nonparametric, zhu2015boosting}. 
The comparison is made by controlling the following quantities: 
\begin{itemize}[leftmargin=*]
    \item[-]Correlation balance: it involves the evaluation of the Pearson correlation between the exposure variable and the entire set of observed covariates;
    \item[-]Marginal balance: authors examine marginal summary statistics for each variable in the dataset, including both the covariates and the exposure variable. This analysis considers both mean and variance measurements;
    \item[-]Effective Sample Size: this summary quantifies the effective number of observations within each sample.
\end{itemize}

By investigating these summaries in both weighted and unweighted distributions, authors can comprehensively assess the performance of the different weighting methods and choose the optimal one for the application. Results of this analysis are shown in Subsection \ref{subsec:results_impact}

\subsubsection{Modeling the impact} \label{subsubsec:impact}

Once the optimal weights are selected, they are used to balance the original dataset and train a regression model on it. Through these analyses, the aim is to achieve a holistic understanding of the influence of commuting time on students' performance in a broader perspective, beyond the correlation between these two variables. Additionally, such correlation is investigated with the application of the same regression model to the original (unbalanced) dataset, allowing both to compare the results from the two training datasets and also to get proper insight into the relationship between GPA and other students' factors.

The regression model chosen for modeling the outcome is a Linear Mixed-effect Model (LMM) \citep{goldstein2011multilevel}. This model allows for the analysis of individual student data while accounting for the variation within the different grouping structures, i.e. the bachelor programs. This is essential as different programs may have distinct characteristics and requirements that can affect students' academic performance (e.g., peer effect, teacher effect). 
Moreover, to gain deeper insights into the impact of students' commuting time on their GPA, a Polynomial Regression analysis is conducted. While simple linear models are known for their simplicity in description and implementation, along with their advantages in terms of interpretation and inference, polynomial linear models take this approach a step further. They expand simple linear models by introducing additional predictors, achieved by raising the original predictors to various powers so that they allow the description of nonlinear relationships between covariates and response \citep{james2013introduction}. Specifically, for this study, a model with commuting time raised to the $D$th-power is chosen, since it has been identified as the optimal power leading to the lowest MSE when applied to a model trained on 70\% of the original observations and subsequently evaluated on the remaining. 

All the available covariates are included in the model: GPA as the response variable, and commuting time and other determinants as predictors. When using the weighted dataset, although it has already been balanced with respect to other covariates than the commuting time, the latter are still included to reduce the bias due to residual imbalance and to gain robustness in the effect estimate \citep{ho2007matching}. Nevertheless, when employing the weighted dataset, the estimates of their coefficients are not interpreted nor examined since may be severely confounded and the final considerations regard the effect of the commuting time only. While the bachelor program is included in the model as a random effect, the remaining determinants are included in the model as fixed effects. Numerical variables appear with an additive relationship to the response. Categorical and binary variables are introduced as dummy variables with respect to a defined baseline, again additive on the response.

The proposed model can be written as: 

\begin{equation} \begin{split}
\label{eq:model_LMM_poly}
y_{il}=& \: \beta_0 + \sum_{j=1}^{J-1} \beta_jx_{jil} + \sum_{d=1}^{D} \beta_{(J-1+d)} a^d_{il} +u_{l} + \epsilon_{il} \\
u_{l} \sim & \: \mathcal{N} \, (0,\sigma^2_{u}) \\ \epsilon_{il} \sim & \: \mathcal{N} \,(0,\sigma^2_{\epsilon})
\end{split} \end{equation}

\noindent where $y_{il}$ denotes the grade point average of student $i$ in program $l$; \mbox{$\bm{\beta}=\{\beta_0,\dots,\beta_{J+D-1}\}$} is the $(J+D)-$dimensional vector of parameters; $x_{jil}$ represents the value of the $j-$th predictor at student's level, in particular $a_{il}$ is the variable of commuting time up to degree $D$; $u_{l}$ is the random effect of the program $l$; $\epsilon_{il}$ is the error; and it is assumed that $\bm{u}$ is independent of $\bm{\epsilon}$.
 
The effect of commuting time on GPA is estimated by looking at the Average Dose-Response Function (ADRF) and the Average Marginal Effect Function (AMEF). The first function links the value of the commuting time to the expected potential GPA under that travel time value across the full sample and estimates the average potential outcome across different levels of exposure: $AMEF(a) = \hat{\mu}(a) = \mathbb{\hat{E}}[Y(a)]$. 60 distinct values of commuting time are examined, ranging from its 5\%-quantile to its 95\%-quantile, and the estimate is computed together with its 90\% confidence intervals. AMEF is also used to characterize the effect of commuting time with the derivative of the ADRF with respect to travel time. When different from zero, it highlights a significant impact of commuting time for its corresponding values. The two estimated curves are compared for the cases of unweighted and weighted datasets, showing the relevance of the balancing procedure for the current application.

\section{Results} \label{sec:results}

\subsection{Commuting time estimation} \label{subsec:results_time_estimation} 

\begin{figure}
\centering
    \subfigure[Leonardo campus]{\resizebox*{6cm}{!}{\includegraphics{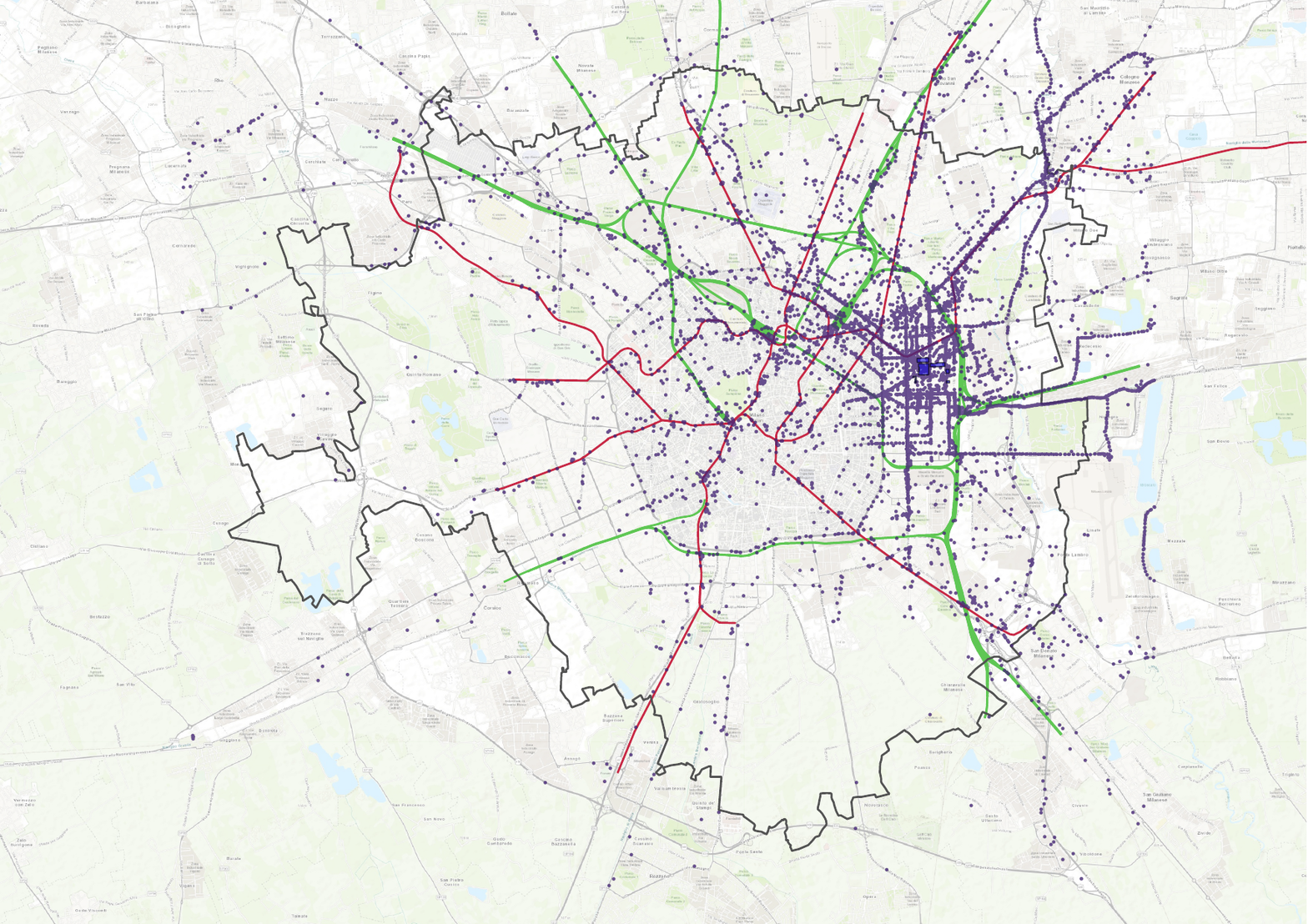}}}\hspace{0.5cm}
    \subfigure[Bovisa campus]{\resizebox*{6cm}{!}{\includegraphics{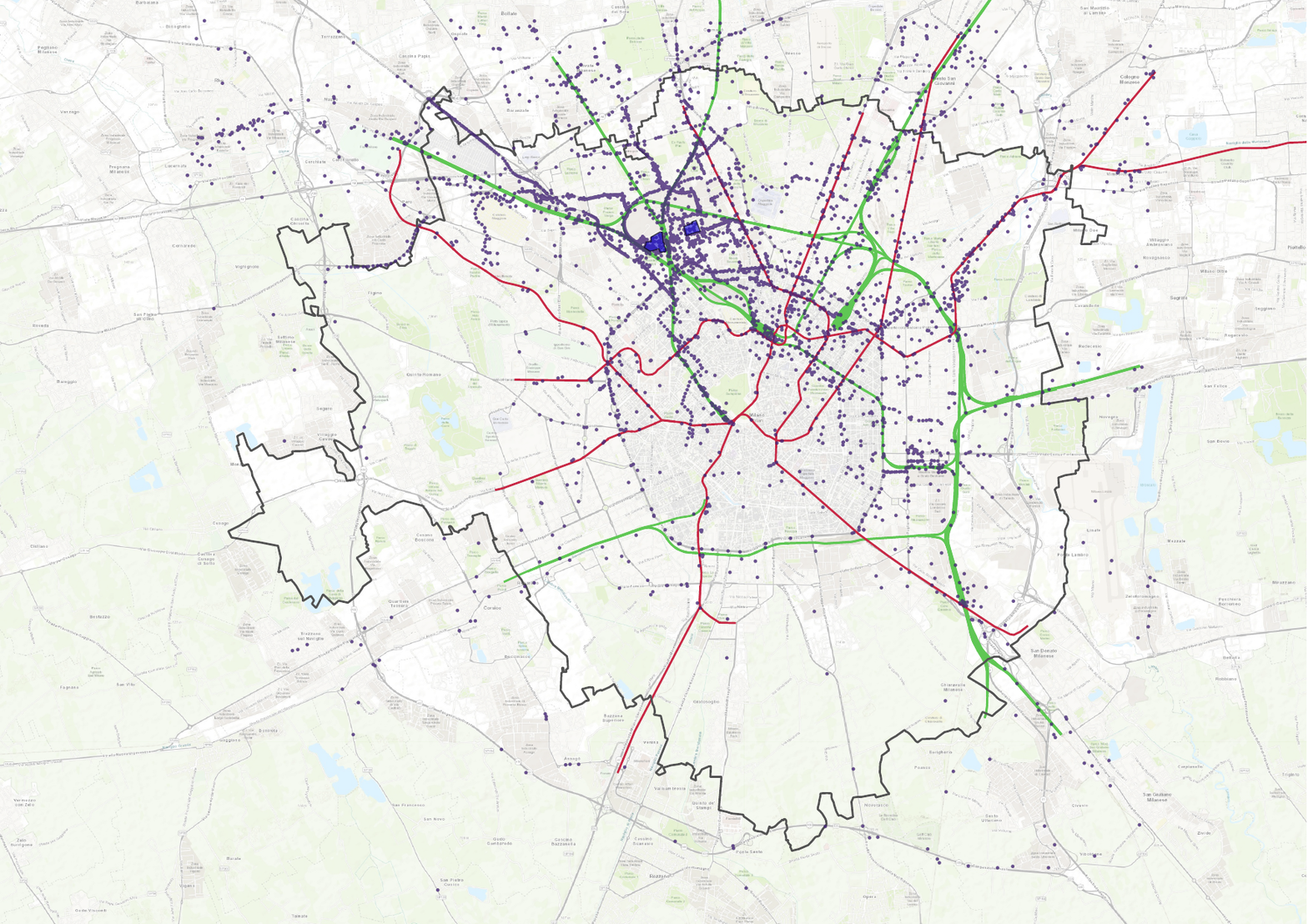}}}
    \caption{Input GPS data within the analyzed area of Milan municipality for the two campus sites.} \label{fig:GPS_Leonardo_Bovisa}
\end{figure}

The commuting time estimation procedure is applied separately to the two campuses of the Politecnico di Milano. The $N_{L}=7,670$ input GPS measurements concerning the Leonardo campus are depicted in Figure \ref{fig:GPS_Leonardo_Bovisa}a, while the $N_B=4,111$ data for the Bovisa site are shown in Figure \ref{fig:GPS_Leonardo_Bovisa}b. They both span within the rectangular box that entirely contains the municipality of Milan, indicated by a gray outline in the figures. The figures also show the location of the university institutions, colored in blue, and the main public transportation lines, with suburban train lines colored in green and subway lines colored in red.

\begin{figure}
\centering
    \subfigure[Leonardo campus]{\resizebox*{6cm}{!}{\includegraphics{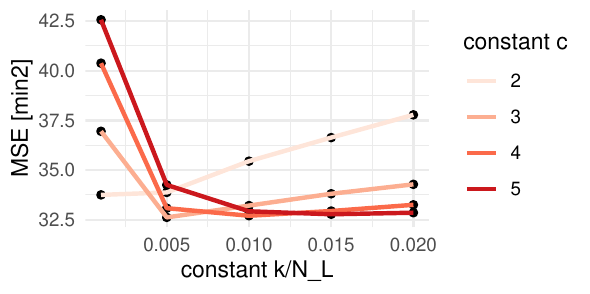}}}
    \hspace{0.5cm}
    \subfigure[Bovisa campus]{\resizebox*{6cm}{!}{\includegraphics{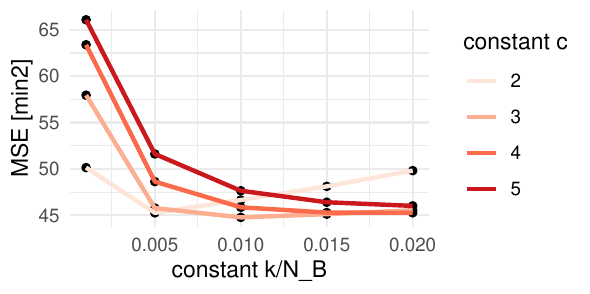}}}
    \caption{MSE of the LOO-CV performed for different parameters $k$ and $c$.} \label{fig:LOOCV_dati_GPS}
\end{figure}

The choice of the hyperparameters $k$ and $c$ for KRE is performed through LOO-CV on the training set. As shown in Figure \ref{fig:LOOCV_dati_GPS}, the MSE is minimized by choosing the combination of $k=0.005N_{L}$ and $c=1/3$ for the Leonardo site (MSE $=32.62\text{ min}^2$), and $k=0.01N_{B}$ and $c=1/3$ for the Bovisa site (MSE $=44.78 \text{ min}^2$). With such parameters, the performance of the methodology is evaluated on a test set of 15\% of the trajectory data. The KRE of commuting time to the Leonardo campus results to be highly accurate, with MSE $= 22.43$ min$^2$ and Mean Absolute Error (MAE) $= 3.24$ min. Concerning the Bovisa site, the performance slightly decreases with MSE = $41.83$ min$^2$ and MAE = $4.62$ min.

\begin{figure}
\centering
    \subfigure[Leonardo campus]{\resizebox*{6cm}{!}{\includegraphics{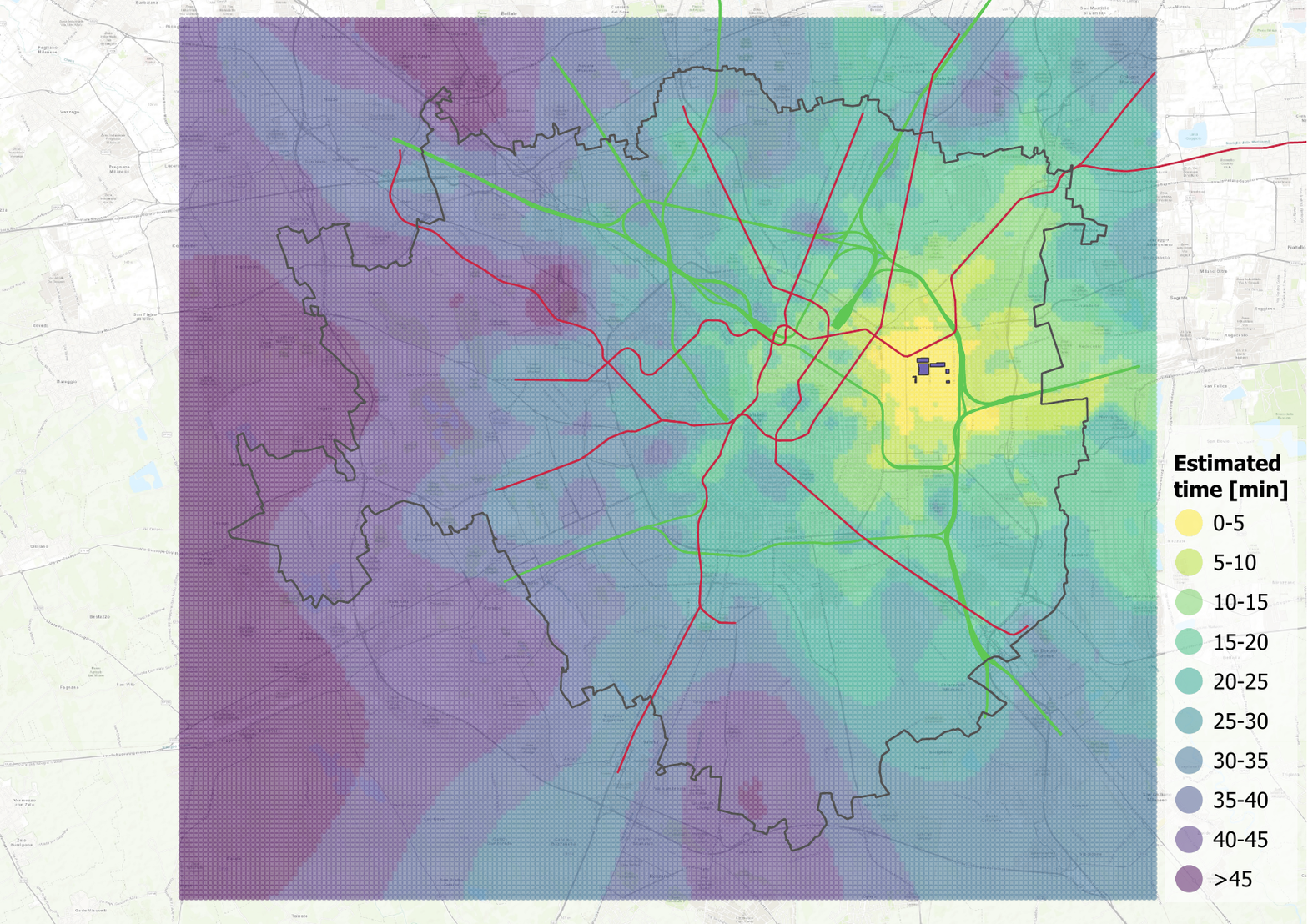}}}
    \hspace{0.5cm}
    \subfigure[Bovisa campus]{\resizebox*{6cm}{!}{\includegraphics{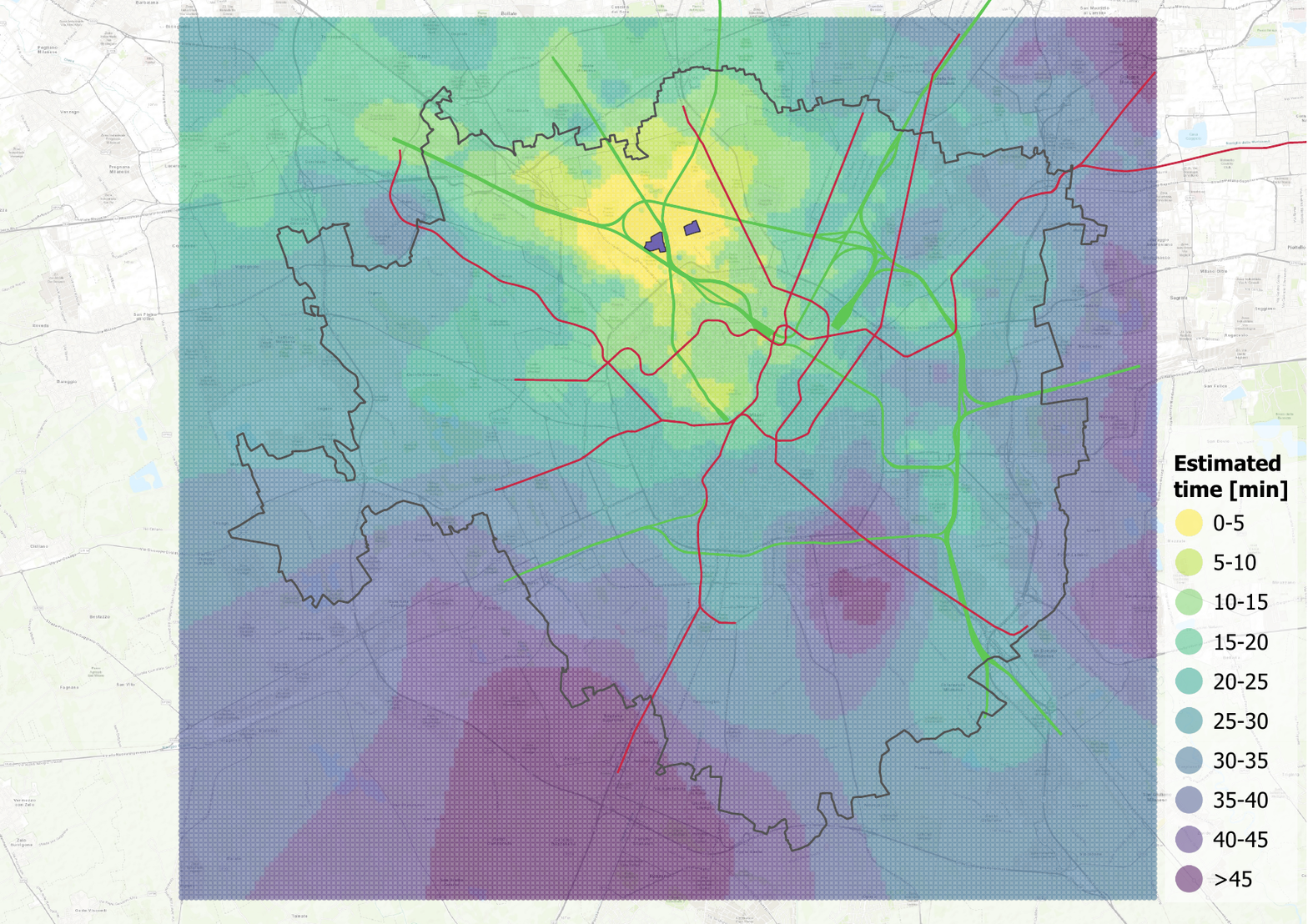}}}
    \caption{Accessibility maps of estimated commuting time, output of KRE, for the two university campuses.}
    \label{fig:mappa_Leonardo_Bovisa}
\end{figure}

The estimated commuting time for each point location on the grid is shown in Figure \ref{fig:mappa_Leonardo_Bovisa} for the accessibility to the Leonardo campus (a) and to the Bovisa campus (b). For both maps, the estimated times generally increase as the distance to the institution increases, as expected. However, it is interesting to point out that the increase does not occur homogeneously in space, but there are directions of anisotropy that follow the main lines of the road network and the public transportation network. For instance, for the Leonardo case, locations in the southwest of the analyzed box, although at a high distance to the university campus, show a low estimated travel time to university when near the subway network. Similarly, for the Bovisa campus, the presence of the highway in the northwest and of the suburban train line in the southeast decreases the estimated commuting time to Bovisa with respect to other locations at the same distance from the university but far from the transportation network.

\subsection{Commuting time and GPA: relationship and impact} \label{subsec:results_impact}

The optimal balancing technique for the study application is identified after comparison with the aim of selecting the best balance for the dataset. Table \ref{tab:correlation_balance}, Table \ref{tab:marginal_balance}, and Table \ref{tab:ess} present the analyzed metrics for balance comparison: correlation balance, marginal balance, and effective sample size, computed for the original dataset and the datasets balanced employing weighting methods.

In the absence of any balancing techniques, the correlation values between travel time and confounding variables span a range from -0.073 to 0.044. This observation underscores the inherent balance within the original dataset. Upon evaluating various balancing approaches, the most significant reduction in correlation is achieved when implementing the methods of EB, npCBPS, and CBPS with multilevel correction.

Each of the employed balancing methods rigorously preserves the integrity of descriptive statistics within the balanced dataset, ensuring that they remain unaltered when compared to their counterparts in the original dataset. This robust consistency reaffirms the reliability and effectiveness of the balancing techniques applied. 

The integrity of the balanced samples is consistently upheld, with their effective size consistently remaining at or above 80\% of the original sample size. It's noteworthy, however, that the application of methods incorporating multilevel correction does lead to slightly smaller effective sample sizes compared to other methods.

Based on these insightful findings, authors are able to make an informed selection of the most suitable balancing method. It's crucial to highlight that all the balancing methods employed can be deemed appropriate for the specific objectives of the project. For the upcoming analyses, authors have opted to implement the Entropy Balancing method with multilevel correction, leveraging its demonstrated effectiveness in optimizing the dataset for further investigation.

\begin{table}[t!]
    \centering
    \tbl{Weights comparison: correlation exposure - confounders\vspace{0.1cm}}{
    \resizebox{\textwidth}{!}{
    \begin{tabular}{l|cccccccccccc}
         & NO & npCBPS & CBPS & GLM & EB & REM & FEM & GLM$_{ml}$ & EB$_{ml}$ & npCBPS$_{ml}$ & CBPS$_{ml}$ \\
         \midrule
         AdmissionAge&0.030 &0.00069 &0.0039 &0.015 &0.0017 &0.018 &0.019 &0.017 &0.0061 &0.0032 & 0.00069 \\
         Gender &-0.073 &0.0084 &0.0038 &0.0046 &0.0040 &0.0090 &0.0098 &0.015 &0.00087 &0.0067 &0.00057 \\
         HighSchoolTrack &0.044 &0.0097 &0.011 &0.0063 &0.011 &0.0055 &0.0017 &-0.0025 &0.010 &0.0062 &0.0096 \\
         HighSchoolGrade &-0.054 &-0.0095 &-0.0029 &0.0018 &-0.0030 &-0.0057 &-0.016 &-0.017 &0.000055 &0.0040 &0.0024 \\
         FamilyIncome &0.039 &-0.00095 &0.0030 &0.015 &0.0027 &0.020 &0.024 &0.026 &0.00029 &-0.0032 &0.00040 \\
         Bachelorprogram&-0.016 &-0.011 &-0.0083 &-0.0085 &-0.0091 &-0.011 &-0.016 &-0.016 &-0.0053 &-0.0055 &-0.0054 \\         
    \end{tabular}}
    }
    \label{tab:correlation_balance}
\end{table}
\begin{table}[t!]
    \centering
    \caption{Weights comparison: descriptive statistics. \\ \textit{Descriptive statistics of the balanced samples compared with those of the original sample size. The mean value and standard deviation are reported for continuous variables while proportion per group for categorical ones \vspace{0.1cm}}}
    
    \resizebox{\textwidth}{!}{
    \begin{tabular}{l|cccccccccccc}
         & NO & npCBPS & CBPS & GLM & EB & REM & FEM & GLM$_{ml}$ & EB$_{ml}$ & npCBPS$_{ml}$ & CBPS$_{ml}$ \\
         \midrule          
         \textbf{AdmissionAge} &0.566 & 0.564 & 0.577 & 0.564 & 0.566 & 0.569 & 0.573 & 0.570 & 0.566 & 0.566 & 0.592 \\
         & (1.106) & (1.122) & (1.190) & (1.113) & (1.124) & (1.112) & (1.105) & (1.095) & (1.090) & (1.080) & (1.206) \\
         \textbf{Gender} \\
         F & 27.0\% & 27.0\% & 26.8\% & 26.7\% & 27.0\% & 27.0\% & 27.3\% & 27.4\% & 27.0\% & 27.0\% & 26.4\% \\
         \textbf{HighSchoolTrack} \\
         Humanistic & 5.72\% & 5.72\% & 5.57\% & 5.56\% & 5.72\% & 5.62\% & 5.75\% & 5.71\% & 5.72\% & 5.73\% & 5.55\% \\
         Scientific & 87.57\% & 87.57\% & 87.63\% & 87.70\% & 87.58\% & 87.70\% & 87.76\% & 87.87\% & 87.57\% & 87.56\% & 88.01\% \\
         Technical & 5.13\% & 5.14\% & 5.14\% & 5.16\% & 5.13\% & 5.11\% & 4.99\% & 4.94\% & 5.13\% & 5.13\% & 4.83\% \\
         \textbf{HighSchoolGrade} & 0.822 & 0.822 & 0.823 & 0.824 & 0.822 & 0.824 & 0.824 & 0.825 & 0.822 & 0.822 & 0.823 \\
         & (0.110) & (0.111) & (0.111) & (0.111) & (0.111) & (0.111) & (0.112) & (0.112) & (0.111) & (0.111) & (0.110) \\
         \textbf{FamilyIncome} \\
         high & 55.42\% & 55.44\% & 55.28\% & 55.25\% & 55.43\% & 55.03\% & 55.19\% & 55.11\% & 55.42\% & 55.44\% & 55.03\% \\
         medium & 18.93\% & 18.90\% & 19.18\% & 19.28\% & 18.93\% & 19.41\% & 19.29\% & 19.41\% & 18.93\% & 18.92\% & 18.87\% \\
         low & 7.50\% & 7.50\% & 7.62\% & 7.51\% & 7.49\% & 7.47\% & 7.44\% & 7.41\% & 7.50\% & 7.50\% & 7.44\% \\
    \end{tabular}}    
\label{tab:marginal_balance}
    
\end{table}

\begin{table}[t!]
    \centering
    \tbl{Weights comparison: effective sample size.\vspace{0.1cm}}{
    \resizebox{\textwidth}{!}{
    \begin{tabular}{l|cccccccccccc}
         & NO & npCBPS & CBPS & GLM & EB & REM & FEM & GLM$_{ml}$ & EB$_{ml}$ & npCBPS$_{ml}$ & CBPS$_{ml}$ \\
         \midrule
         ESS\hspace{0.5cm} & 507 &492 &494 &494 &495 &488 &471 &465 &456 &432 &405 \\ 
    \end{tabular}}
    }
    \label{tab:ess}
\end{table}

Finally, the relationship and the impact of commuting time on GPA are investigated through the multilevel model in Model \ref{eq:model_LMM_poly}. The model is firstly trained on the original unweighted dataset and secondly on the balanced weighted dataset. Analyzing the unweighted dataset allows for both delving into the determinants influencing students' school performance and for obtaining baseline results on the time-GPA relationship, to be further compared with the time-GPA impact. Parallelly, the analysis of the balanced dataset leads to the identification of the causal impact of interest.

\begin{figure}[b!]
    \centering
    \includegraphics[width=5cm]{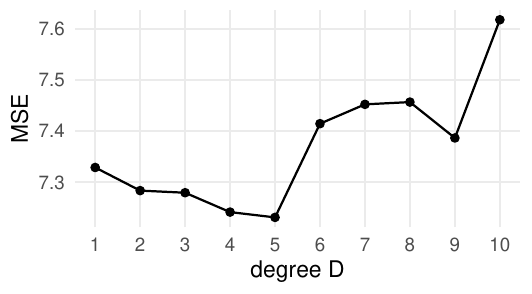}
    \caption{MSE (y-axis) evaluated on the test set when models of different degrees (x-axis) are trained.}
    \label{fig:mse_polynomial_degree}
\end{figure}

Regarding the unweighted dataset, at first, the optimal degree for the variable representing students' commuting time estimate is determined via MSE minimization. Figure \ref{fig:mse_polynomial_degree} shows the trend of the MSE on the test set when the polynomial in the model is defined for different degrees $D\in\{1, \dots, 10\}$ on a training set of 70\% individuals. The model that minimizes the MSE is the one of degree $D=5$.

\begin{figure}
\centering
    \subfigure[Estimated fixed coefficients]{\resizebox*{6cm}{!}{\includegraphics{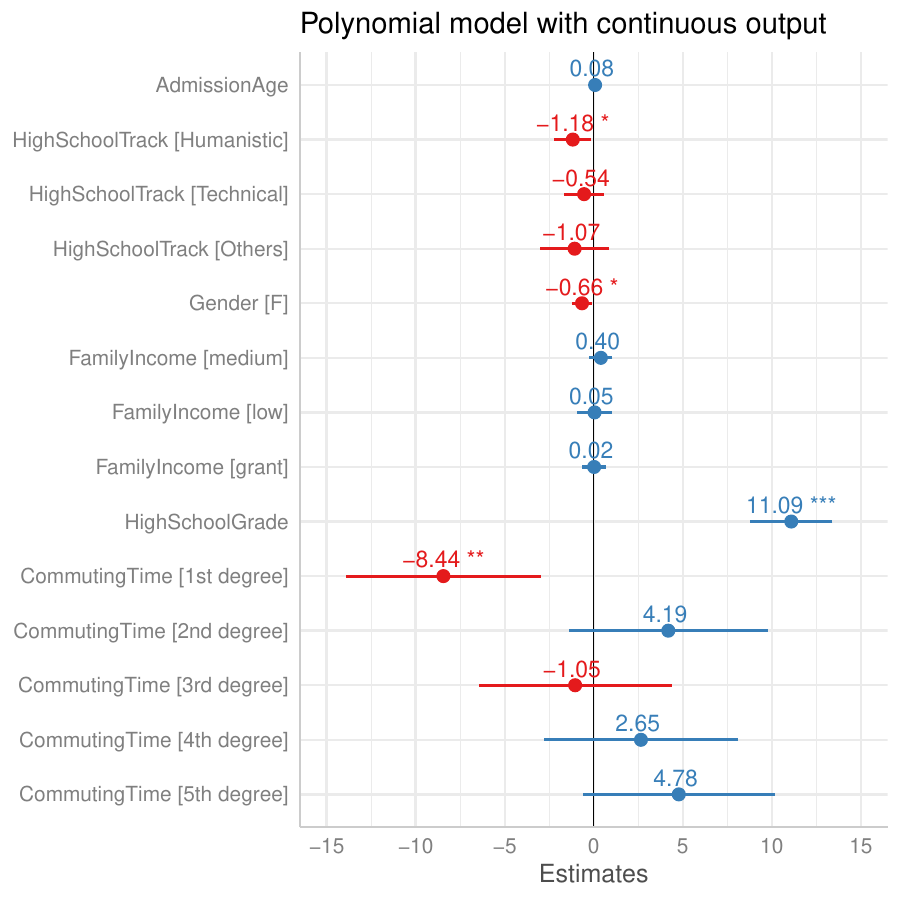}}}
    \hspace{1cm}
    \subfigure[Random effects]{\resizebox*{6cm}{!}{\includegraphics{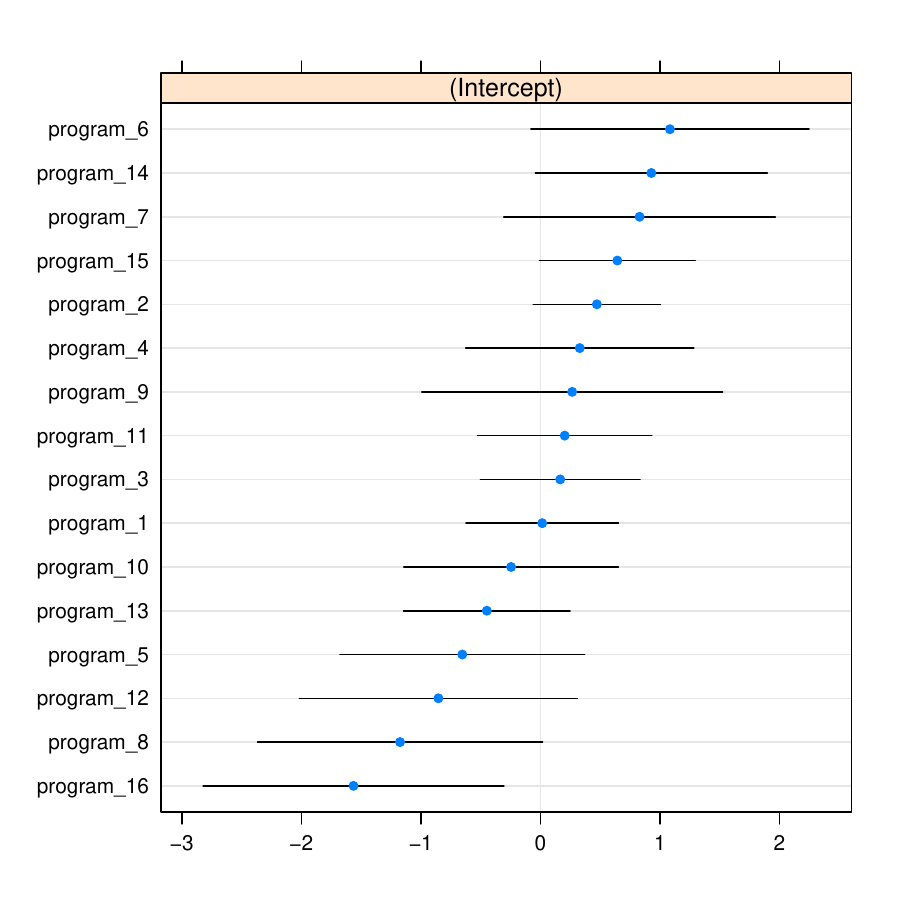}}}
    \caption{Estimated coefficients of the polynomial model with continuous output (Mod. \ref{eq:model_LMM_poly}).}
    \label{fig:mod_polynomial}
\end{figure}

\begin{figure}
\centering
    \subfigure[Average Dose-Response Function]{\resizebox*{6.5cm}{!}{\includegraphics{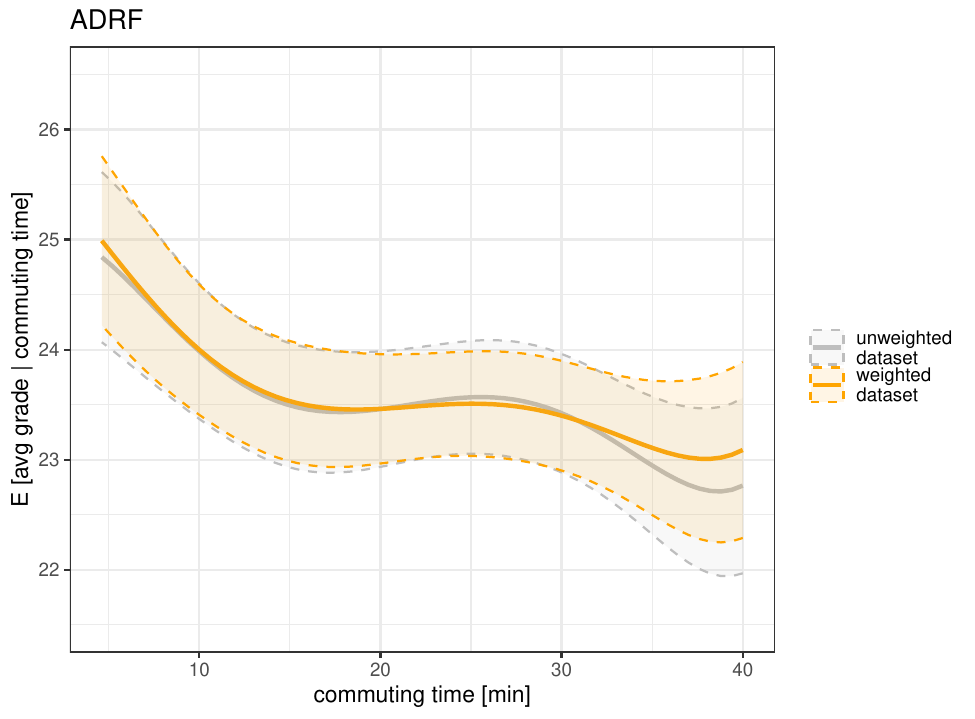}}}
    \hspace{0.5cm}
    \subfigure[Average Marginal-Effect Function]{\resizebox*{6.5cm}{!}{\includegraphics{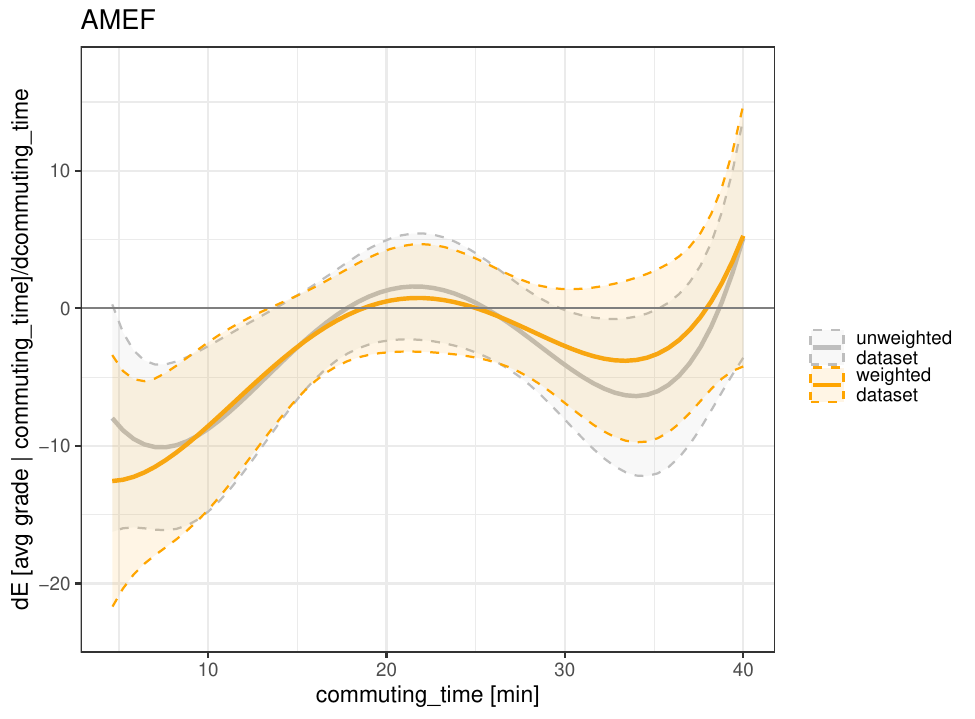}}}
    \caption{(a) ADRF of commuting time on the potential outcome, with confidence bands of 90\%; (b) AMEF relating commuting time to the derivative of the ADRF, with confidence bands of 90\%. In the case of the unbalanced dataset lines are shown in grey, while for the balanced dataset through Entropy weights they are depicted in orange.}
    \label{fig:effect_Weight}
\end{figure}

The performance of the model is evaluated through LOO-CV on the complete dataset and returns a Root Mean Square error (RMSE) of $2.785$. The significant covariates are the gender of the student, with a negative impact of -0.66 for being a female, the previous achievements in high school, increasing the average GPA of 0.1109 for every point in the high school grade, and the high school track, penalizing students from humanistic high schools than scientific ones. Figure \ref{fig:mod_polynomial} collects these results. Interestingly, the commuting time remains significant for explaining the variability of the GPA. The AMEF of commuting time on the model outcome, shown in Figure \ref{fig:effect_Weight} with a grey line, is significantly different from zero when the commuting time is less than 15 minutes and for higher values between 30 and 40 minutes, while there is no impact of commuting time for journeys of medium duration between 15 minutes and 30 minutes. 

The marginal effect of commuting time on students' performances is computed for the balanced weighted dataset and analyzed. Both from the ADRC and the AMEF, shown in Figure \ref{fig:effect_Weight} with an orange color, the impact of commuting times between 5 and almost 15 minutes is significant on students' performances at the 90\% confidence level, almost-linearly decreasing the average grade from 25/30 to 23.5/30. For larger times, however, travel time does not have a statistically significant effect on academic performance, being the AMEF non-significantly different from zero for travel times larger than 15 minutes. These results partially align with the ones of the analysis of the unbalanced dataset, showing the same general behavior of the curves, as in Figure \ref{fig:effect_Weight} colored in grey. It is interesting to notice, however, that the marginal effect of commuting time would have resulted significant at 90\% also for the range of values around 35 minutes that, thanks to the balancing procedure, is contrarily spotted to have a non-significant effect.

\section{Conclusions} \label{sec:conclusions}

This study implies a two-step procedure to evaluate the effect that commuting time has on the academic performance of university students focusing on the metropolitan area of Milan and the students of the Politecnico di Milano. The first step of the analysis uses machine learning algorithms and GPS data from anonymous users to reconstruct accessibility maps of the two university campuses. The latter leads to the estimation of commuting time for each student, whose information is used in the next step. The second step involves an analysis of the determinants of student performance in terms of GPA at the end of the first semester. This analysis reveals a negative dependency with statistical significance of travel time on student performance, which is why this study aims to go deeper and investigate the causal relationship of these factors. A search for a suitable balancing method follows and the selected one turns out to be the Entropy Balancing method with multilevel correction. Finally, in the causal inference setting the previous conclusions are confirmed.

The innovation and importance of this work lie in the application of the current analytical and methodological framework, but also in the implications of this study. It aims to analyze a key determinant of students' academic performance, i.e. commuting time, that has not been extensively and thoroughly investigated in the literature to date. In the first phase of the study, an innovative data source such as GPS measurements is used and a new methodology specific to this context is adopted. The results show that the proposed data-driven approach is crucial to accurately represent the mobility dynamics of a metropolitan city like Milan, where physical distance is not a reliable measure of accessibility, and where the transportation network is influenced by service availability and traffic conditions. In the second phase of the study, on the other hand, a causal relationship is examined, using balancing methods drawn from the recent literature and once again adopting an innovative methodological approach for this field. To the best of the authors' knowledge, this is the first time that the causal relationship between commuting time and academic performance is highlighted and statistically tested using recent techniques of causal inference like balancing weighting methods. 

The implications of this research are highly relevant for the improvement and support of university students. One of the key findings underscores the crucial importance of positioning university campuses in well-connected and easily accessible areas within the city. This discovery suggests that, when planning new campuses, prioritizing strategic locations to ensure easy accessibility for students from various parts of the city should be paramount. This initial step can significantly enhance the learning experience of students and contribute to optimize their academic performance.
Simultaneously, it is essential to develop and support an efficient and well-structured urban transportation system designed to make commuting to existing campuses as seamless as possible. The goal is to reduce commuting times for students, allowing them to allocate more time to their studies and academic activities.
Another significant implication of this study is the need to strategically position student residences in relation to the campuses. If housing facilities are situated in close proximity to the study locations, students will benefit from substantial reductions in commuting times to attend classes. This not only enhances the quality of students' lives but can also contribute to increased participation in academic activities and engagement in university life.
Finally, it is important to highlight that the findings of this study may have relevance in urban contexts other than the city of Milan. The methodology used and the implications drawn could be replicated and adapted to other cities, thereby contributing to the optimization of the learning environment for university students in settings similar to the Milan municipality. In line with this, future developments could focus on assessing the reproducibility of this study in similar contexts, encompassing not only other Italian metropolitan cities but also European ones.
Moreover, to enhance its value and appeal, it would be highly advantageous to broaden the scope of analysis beyond the central area of Milan. This entails considering the surrounding areas (e.g., the province of Milan) and incorporating the perspective of commuting students.

\section*{Acknowledgement}
The authors thank Cuebiq Inc. for sharing the GPS dataset used in this work.
\\
The authors acknowledge the support by MUR, grant Dipartimento di Eccellenza 2023-2027. 
\\
Arianna Burzacchi's work has been further supported by the Next Generation EU Programme REACT-EU through the PON Ph.D. scholarship “Development of innovative Eulerian privacy-preserving data analysis tools for designing more sustainable and climate-friendly human mobility services and infrastructures from high-resolution location data”.

\newpage

\bibliographystyle{tfcad}
\bibliography{MAIN_PAPER}

\newpage
\appendix
\addcontentsline{toc}{section}{Appendices}

\section{Trajectory estimation}
\addcontentsline{toc}{subsection}{Trajectory estimation}
\label{appendix1}

The dataset provided by Cuebiq is extensive and has a wide spatiotemporal coverage. When limited to the metropolitan area of Milan and working days, it contains 26,397,613 GPS signals from 165,582 users’ smartphones, describing their position in time both during spatial movements and stopovers. Among all this information, however, the current research is interested in obtaining measurements of time-to-university, i.e., just the ones related to journeys (and not stops) directed toward the university site. The wide original dataset is hence reduced and further elaborated to extract this valuable information and make it usable for accessibility map constructions.

The dataset is composed by GPS signals $P_{i}=(x_{1i}, x_{2i}, t_i)$, representing the $i$-th GPS sample measured at time $t_i$ at UTM coordinates $x_{1i}$ and $x_{2i}$. The observation $P_{i}$ describes either the situation of a moving subject (e.g., the user is traveling from one place to another), the situation of a temporarily non-moving subject that will restart its journey (e.g., the user is waiting for the bus), or the situation of a permanently non-moving subject (e.g., the user is at home). The first two cases are associated with movements and their GPS signals are called moving points, while data associated with non-moving status are called stopping points. 

As a first step, it is needed to elaborate and filter the original dataset of GPS signals $\{P_{i}\}_{i}$ to create a new version where stopping points are discarded and the remaining signals are grouped together when they describe the same spatial movement of the same user. The group of signals is the so-called journey, also known as trip or trajectory. A journey describes the spatial movements of a single user through a sequence of GPS data points, ordered with respect to the timestamp, which starts from a moving point and ends at a stopping point.
Following the same notation as before, the $k$-th journey is characterized by the sequence of $n_k$ points from the first one after a set of stopping points $P_{m^k_{1}}$ until the occurrence of a new stopping point $P_{ m^k_{n_k}}$, and can be written as $M_{k} = \{P_{m_1^k},\dots, P_{m_{n_k}^k}\}$.

In order to assess the identification of trajectory, it is necessary to define standards for the detection of stopping points. These guidelines originate from the general and intuitive notion of stopping point, and are then formalized with the application of precise quantitative criteria.
Consider two consecutive GPS signals of the same user $P_a = (x_{1a}, x_{2a}, t_a)$ and $P_b = (x_{1b}, x_{2b}, t_b)$, with $t_b > t_a$, and define the distance traveled $\Delta x = \sqrt{(x_{1b}-x_{1a})^2 + (x_{2b}-x_{2a})^2}$, the elapsed time $\Delta t = t_b - t_a$, and the average velocity $v = \Delta x / \Delta t$ between the two measurements. 
The selected identification criteria for stopping points set global constraints on distance, time, and velocity, applicable for all users, and additional device type-based constraints. Global criteria assess the measurements between consecutive instances and highlight the stability of locations over time through the introduction of thresholds on distance ($\Delta x = 0 \text{ m}$), time ($\Delta t > 60\text{ min}$), and velocity ($v<0.7\text{ m/s}$) to identify stopping points. Additional device type-based constraints are introduced to account for the data-collecting procedure, which differs according to the user’s smartphone operating system: iOS-based devices sample new measurements after movements of about 50 meters from the previously recorded position, while Android devices after approximately 5 minutes. Hence additional constraints are added on time and velocity ($\Delta t<15\text{ min}$ and $v<1\text{ m/s}$) in the Android case and on distance and velocity ($\Delta x<100 \text{ m}$ and $v<1 \text{ m/s}$) in the iOS case.  

Once the stopping points are identified and the trajectories are found, the full dataset of $\{M_{k}\}_{k}$ is reduced to consider just the journeys that end in the proximity of the two university campuses: if the final stopping point of a journey is at most 250 meters away from the site, then the trajectory is kept; otherwise, it is discarded. The reduction is needed to get a GPS dataset representative of the mobility pattern to any point in Milan to the university site, which mimics the behavior of a student traveling for attending lectures. Further filtering is applied to ensure more robustness in the measurements, removing journeys with less than 6 observations, journeys with at least one signal of low accuracy ($acc\geq1,500 \text{ m}$), and self-loop journeys from and to the same university campus. Upon the conclusion of this phase, the dataset pertaining to the Leonardo campus consists of 7,670 GPS measurements of 881 trajectories from 448 unique Cuebiq users. Trajectories are composed of 8.6 signals on average and have a mean length of 7.8 kilometers, ranging from 300 meters to a maximum of 33 kilometers. More than 62\% of Cuebiq users are associated with only one trajectory, almost 18\% with two trajectories, and the few remaining can reach 10 trajectories. Similarly, in relation to the Bovisa campus, there are 4,111 GPS measurements associated with 498 trajectories from 295 distinct Cuebiq users. The dataset is almost half the size of the previous one. On average, with respect to the case of Leonardo campus, trajectories result longer in terms of distance traveled (8.7 kilometers on average) and shorter in terms of number of signals (7.92 on average) and user (1.6 trajectories per user on average). All this information is summarized in Table \ref{tab:trajectories}.

Finally, the commuting time to university of each GPS measurement is computed as the amount of time spent from its measurement to the end of the trajectory. Let us consider the $k$-th journey $M_k$. The commuting time to university $y_{m^k}$ associated to the point $P_{m^k} = (x_{1m^k}, x_{2m^{k}}, t_{m^k})$ for any $m=m_1^k, \dots, m_{n_k}^k$ is computed as $y_{m^k} = t_{m_{n_k}^k}-t_{m^k}$.

\begin{table}[h]
    \centering
    \tbl{Descriptive statistics of the GPS dataset after elaboration and filtering for both the campus of Leonardo and Bovisa} {
    \footnotesize
    \begin{tabular}{l c  c}
    \toprule
        Descriptive statistics & Leonardo campus & Bovisa campus \\
        \midrule
        n. signals & 7,670 & 4,111 \\ &&\\
        n. trajectories & 881 & 498 \\ &&\\
        n. signals per trajectory & [6; 85] & [6; 55]\\
        & mean=8.6 & mean=7.92 \\ &&\\
        distance per trajectory [km] \hspace{0.5cm} & [0.3; 33.0] & [0.6; 36.5] \\
        & mean=7.8 & mean=8.7\\ &&\\
        n. unique anonymous users & 448 & 295 \\ &&\\
        n. signals per anonymous user & [6; 183] & [6; 149] \\ 
        & mean=15.7& mean= 13.9\\ &&\\
        n. trajectories per anonymous user & [1; 10] & [1; 8] \\
        & mean=1.8 & mean=1.6\\
        \bottomrule
    \end{tabular}

    }    \label{tab:trajectories}
\end{table}

\section{Qualitative analysis on the representativeness of the students' sample}
\addcontentsline{toc}{subsection}{Representativeness of the sample of students}
\label{appendix2}

\begin{table}[b!]
    \centering
    \tbl{Comparison of samples descriptive statistics} {
    \footnotesize
    \begin{tabular}{l c  c c c}
    \toprule
        \textbf{Variable's name} & \textbf{Group1\textsuperscript{a} N=6166} & \textbf{Group2\textsuperscript{b} N=4692}& \textbf{Group3\textsuperscript{c} N=669}& \textbf{Group4\textsuperscript{d} N=507}\\
        \midrule
        \textbf{Gender} \\
        Male& 4596 (74.5\%)&3524 (75.1\%)&490 (73.2\%)&370 (73.0\%)\\
        Female& 1570 (25.5\%)&1168 (24.9\%)&179 (26.8\%)&137 (27.0\%)\\ &&&&\\
        \textbf{Admission age}\\
        Range &[-1;37]&[-1;30]&[-1;14]&[-1;14]\\
        Mean & 1.022&0.66&0.67&0.57\\
        SD&2.69&1.39&1.23&1.11\\&&&&\\
        \textbf{Family income}\\ 
        High&2567 (41.6\%)&1809 (38.6\%)&380 (56.8\%)&281 (55.4\%)\\
        Middle &1288 (20.9\%)&1091 (23.3\%)&115 (17.2\%)&96 (18.9\%)\\
        Low& 686 (11.1\%)&554 (11.8\%)&48 (7.2\%)&38 (7.5\%)\\
        Grant&1625 (26.4\%)&1238 (26.4\%)&126 (18.8\%)&92 (19.2\%)\\&&&&\\
        \textbf{High school grade}\\
        Range& [0.6;1]& [0.6;1]& [0.6;1]& [0.6;1]\\
        Mean& 0.85&0.87&0.81&0.82\\
        SD&0.12&0.11&0.12&0.11\\&&&&\\
        \textbf{High school track}\\
        Humanistic& 300 (4.9\%)&203 (4.3\%) &44 (6.6\%)&29 (5.7\%)\\
        Scientific& 4864 (78.9\%)&3848 (82.0\%)& 565 (84.5\%)&444 (87.6\%)\\
        Technical& 829 (13.5\%)&552 (11.8\%)& 47 (7.0\%)&26 (5.1\%)\\
        Other& 173 (2.8\%)&89 (1.9\%)&13 (1.9\%)&8 (1.6\%)\\&&&&\\
        \textbf{Commuting time}\\
        Range&&&[0;0.76]&[0;0.76]\\
        Mean&&&0.38&0.38\\
        SD&&&0.17&0.17\\&&&&\\
        \textbf{GPA}\\
        Range&&[0;12]&&[0;12]\\
        Mean&&5.9&&5.6\\
        SD&&3.02&&3.04\\
        \bottomrule
    \end{tabular}
    }    
    \tabnote{ \raggedright{
     \textsuperscript{a}First-year Engineering Bachelor students in a.y. 2019/20;
     \textsuperscript{b}First-year Engineering Bachelor students in a.y. 2019/20 who pass at least one exam;
     \textsuperscript{c}First-year Engineering Bachelor students in a.y. 2019/20 resident in Milan;
     \textsuperscript{d}First-year Engineering Bachelor students in a.y. 2019/20 resident in Milan who pass at least one exam.
    }}
    \label{tab:statistics_comparison}
\end{table}

\begin{figure}
\centering
    \subfigure[Gender]{\resizebox*{6cm}{!}{\includegraphics{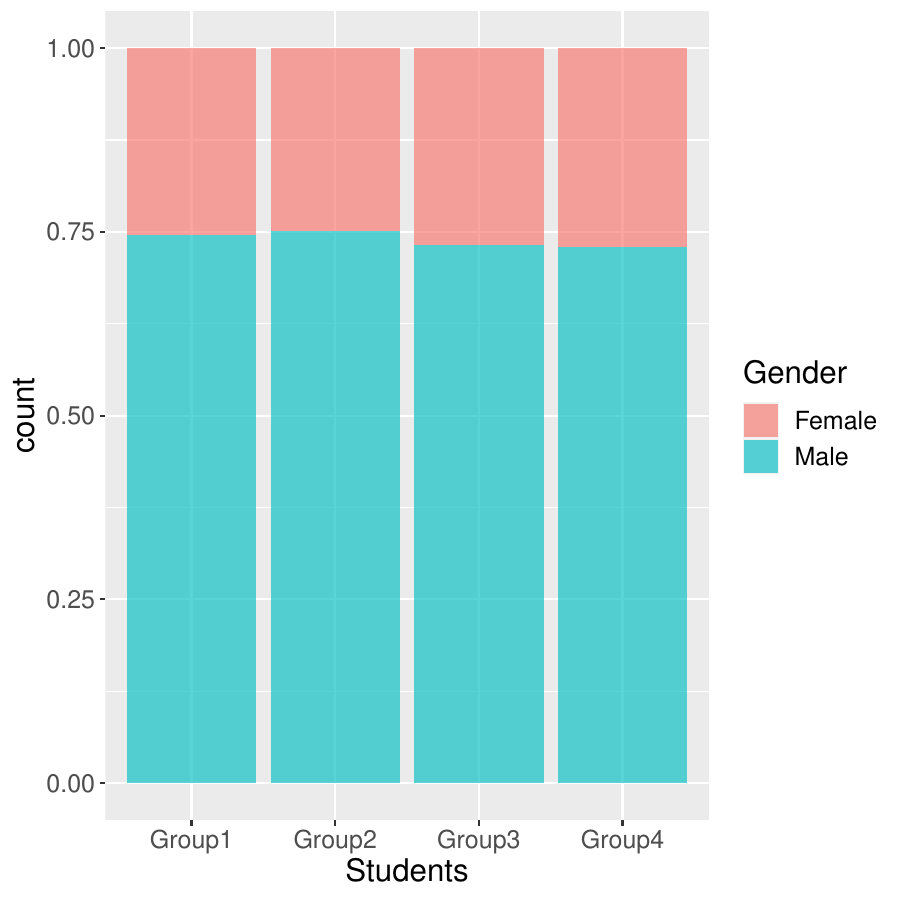}}}\hspace{0.5cm}
    \subfigure[Admission age]{\resizebox*{6cm}{!}{\includegraphics{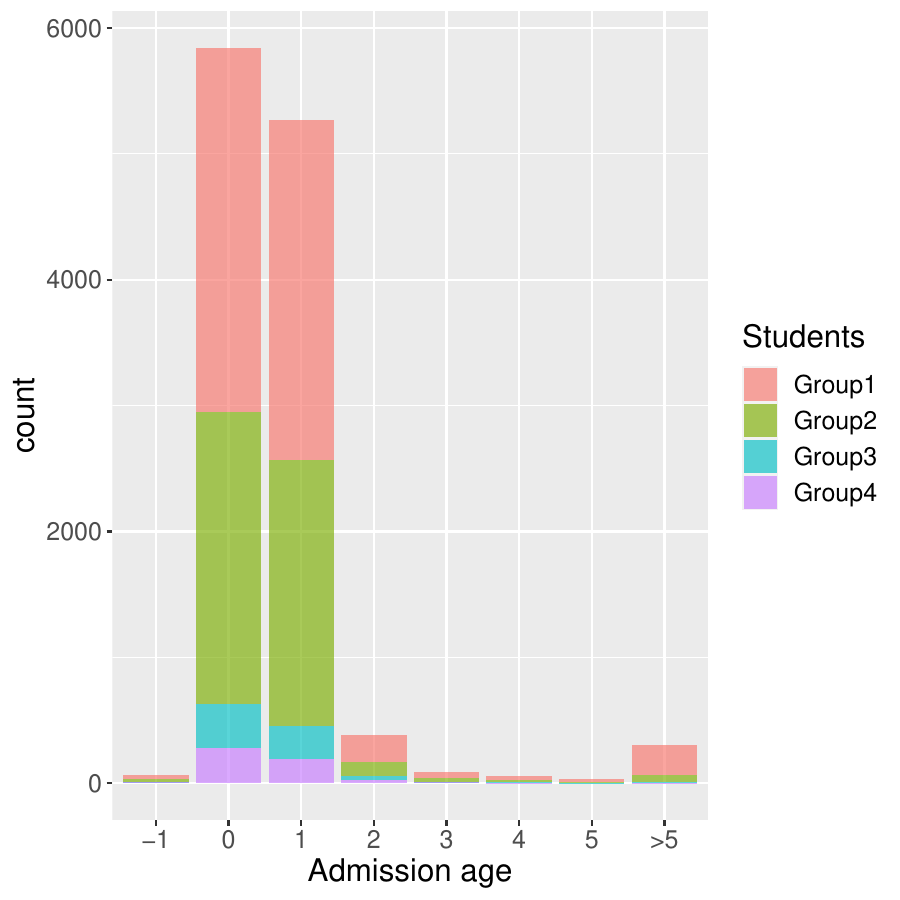}}}
    \hspace{0.5cm}
    \subfigure[Family income]{\resizebox*{6cm}{!}{\includegraphics{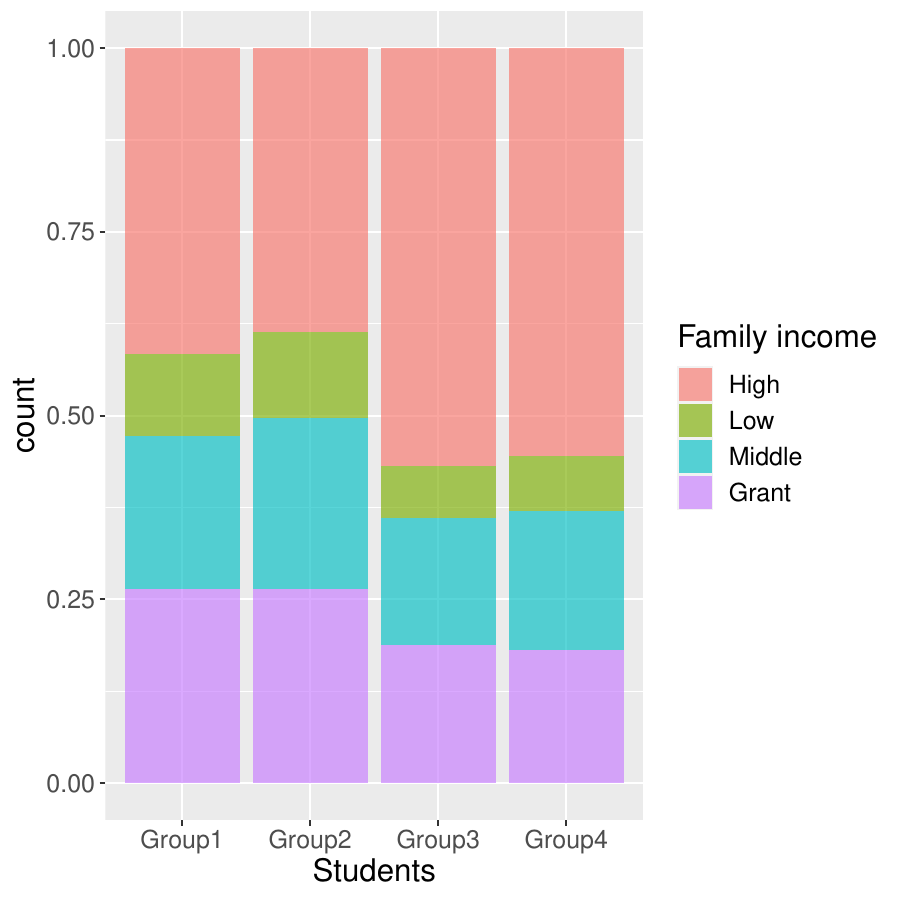}}}
    \hspace{0.5cm}
    \subfigure[High school grade]{\resizebox*{6cm}{!}{\includegraphics{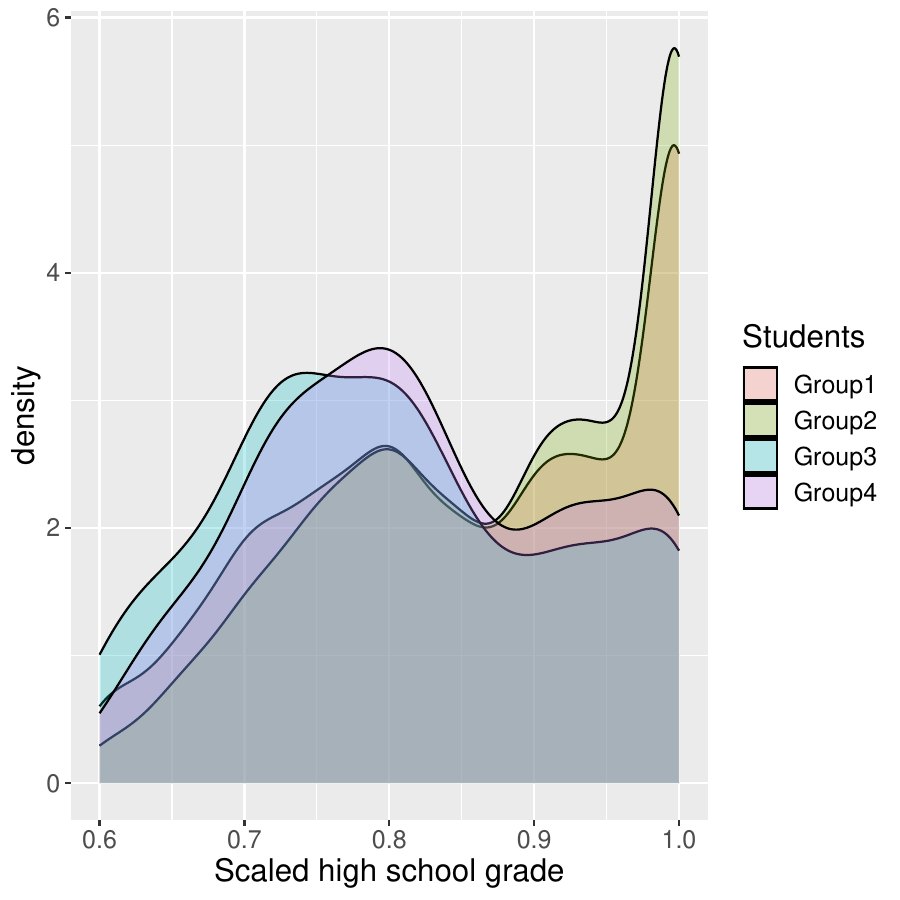}}}
    \hspace{0.5cm}
    \subfigure[High school track]{\resizebox*{6cm}{!}{\includegraphics{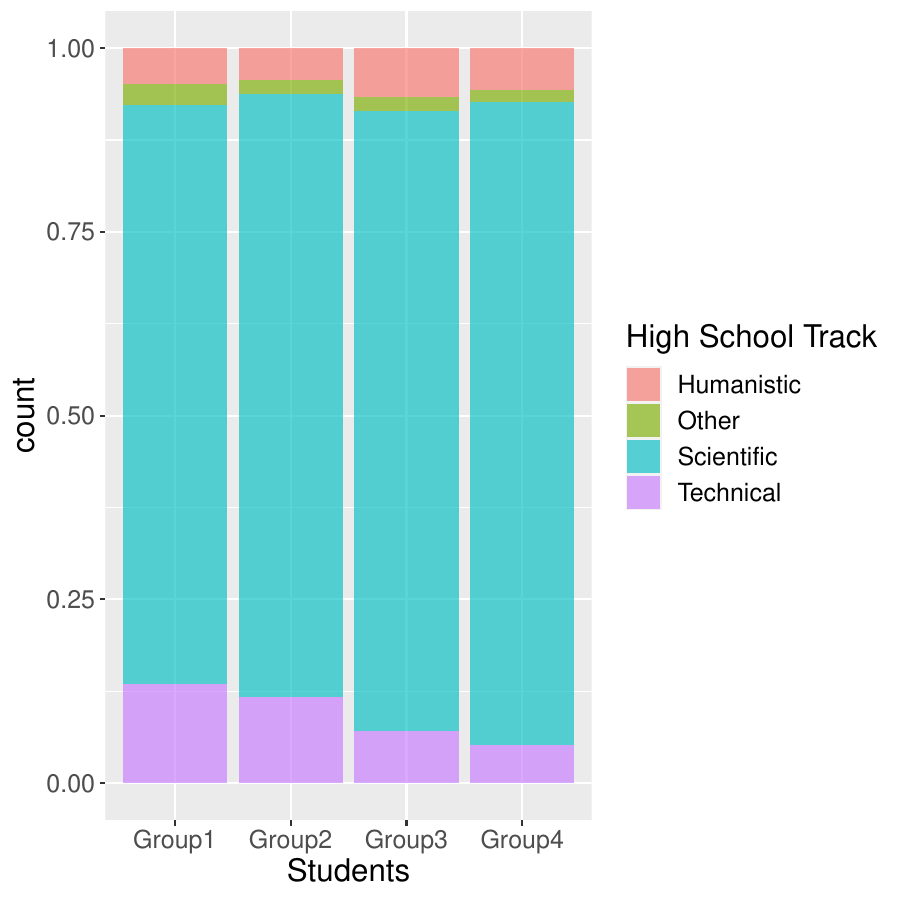}}}
    \hspace{0.5cm}
    \subfigure[GPA]{\resizebox*{6cm}{!}{\includegraphics{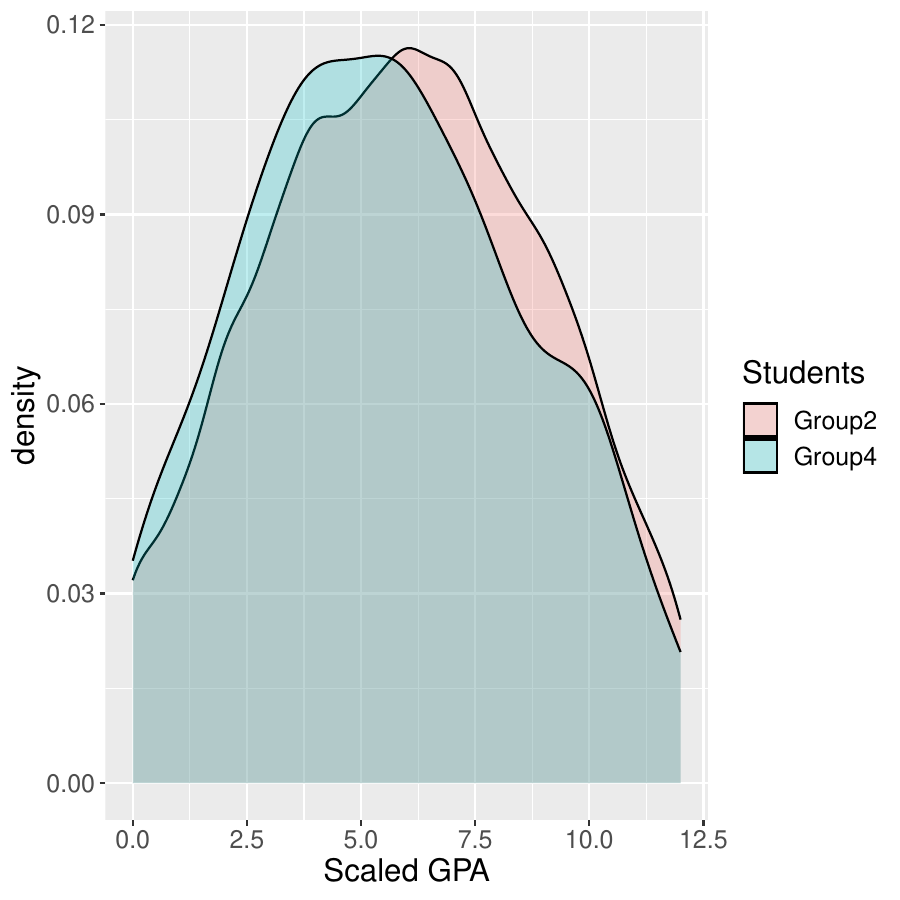}}}
    \tabnote{ \raggedright{
     Group 1: first-year Engineering Bachelor students in a.y. 2019/20; 
     Group 2: first-year Engineering Bachelor students in a.y. 2019/20 who pass at least one exam;
     Group 3: first-year Engineering Bachelor students in a.y. 2019/20 resident in Milan;
     Group 4: first-year Engineering Bachelor students in a.y. 2019/20 resident in Milan who pass at least one exam.
    }}
    \caption{Comparison of samples distributions} 
    \label{fig:comparison_distribution}
\end{figure}

The sample students selected for the analyses are a small portion of the entire population of first-year engineering students at Politecnico di Milano. Indeed, there are considered only students who passed at least one exam and whose residence location is in the Milan metropolitan area. This fact may limit the interpretation of the results to the only study case, thus preventing their generalization to the entire student body. Hence, this section aims to demonstrate the representativeness of the selected sample and to assess the degree to which the findings can be extended. Firstly, a comparison between the groups of students is carried out by looking at the summary statistics of the variables. Then, focusing on the restriction of students living in Milan, a regression model is used to investigate the relationship between the indicator of passing and not at least one exam and the commuting time.

More formally, four distinct groups of students are defined. Group 1 represents the entire population of first-year students enrolled in the bachelor engineering program at the Politecnico di Milano during the first semester of the academic year 2019/20, with a total of 6166 students. Group 2, comprising 4692 students, includes those students in Group 1 who have successfully completed at least one exam by the end of the first semester. 
The sample is then restricted to students residing in Milan. Group 3 consists of those 669 individuals of Group 1 who live in Milan (i.e., first-year bachelor students in engineering at Politecnico di Milano living in Milan). Parallelly, Group 4 encompasses 507 students of Group 2 who are residing in Milan (i.e., first-year bachelor students in engineering at Politecnico di Milano living in Milan and passed at least one exam by the end of the first semester), and composes the sample used in the analyses. Note that for students in Groups 2 and 4 it is possible to compute the GPA, while the remaining students are not associated with any mark. 

Table \ref{tab:statistics_comparison} shows the main descriptive statistics. With the same objective, Figure \ref{fig:comparison_distribution} illustrates the distributions of the analyzed variables in the four samples. 
No major differences emerge with regard to gender and age of admission between different groups. However, disparities are noted in terms of household income, with apparently higher income among Milan residents. In addition, the latter appear to have taken more science and humanities and less technical courses during high school, and have obtained lower final grades in high school. Finally, in terms of GPA, although similar mean values occur in the two samples, the GPA distribution seems slightly shifted to smaller values for Group 4 than for Group 2.

Finally, the study employs a Generalized Linear Mixed-effect Model (GLMM) to examine the association between the variable indicating whether a student passes at least one exam at the end of the first semester or not and the commuting time, restricting the analysis to students who live in Milan. The goal of the analysis is to highlight the relevance of commuting time in distinguishing between students passing or not an exam, i.e., belonging to Group 3 or Group 4. 
For each student $i$, $i=1,\dots,n_{l}$ in the program $l$, $l=1,\dots,L$, $n=\sum_{l}n_{l}$, the model can be written as: 
    \begin{equation} \begin{split}
    y_{il} \sim & \: Be(p_{il}) \hspace{2pt} \\
    logit(p_{il})= &\: \beta_0+\sum_{j=1}^{J-1}\beta_j x_{jil}+\beta_J a_{il}+u_{l} \\
    u_{l} \sim & \: N(0,\sigma^2_{u_2})
    \end{split}
    \label{eq:model_binary}
    \end{equation}
    
where $y_{il}$ is the output variable of student $i$ (1 if they pass at least one exam), in program $l$; $ \bm{\beta}=\{\beta_0,\dots,\beta_J\}$ is the $(J+1)$-dimensional vector of parameters; $x_{jil}$ is the value of the $j$-th predictor at student's level;  $a_{il}$ represents the commuting time; $u_{l}$ is the random effect of the program $l$.

\begin{figure}
\centering
    \subfigure[Odds Ratio]{\resizebox*{6cm}{!}{\includegraphics{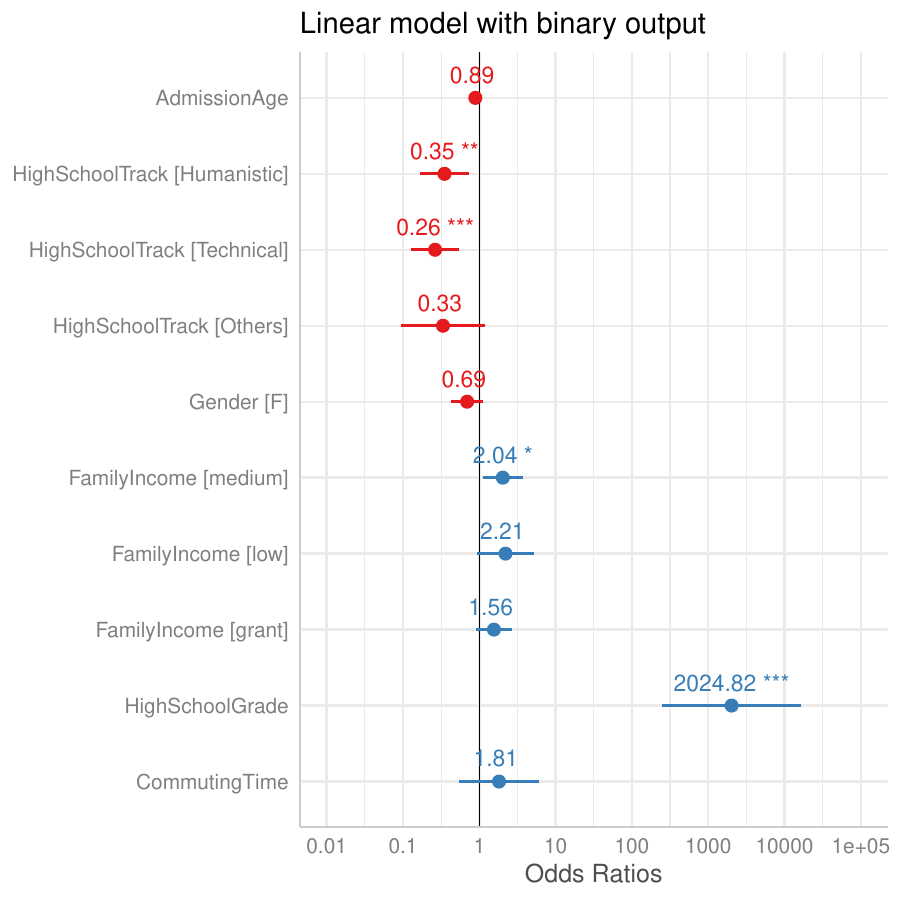}}}\hspace{0.5cm}
    \subfigure[Random effects]{\resizebox*{6cm}{!}{\includegraphics{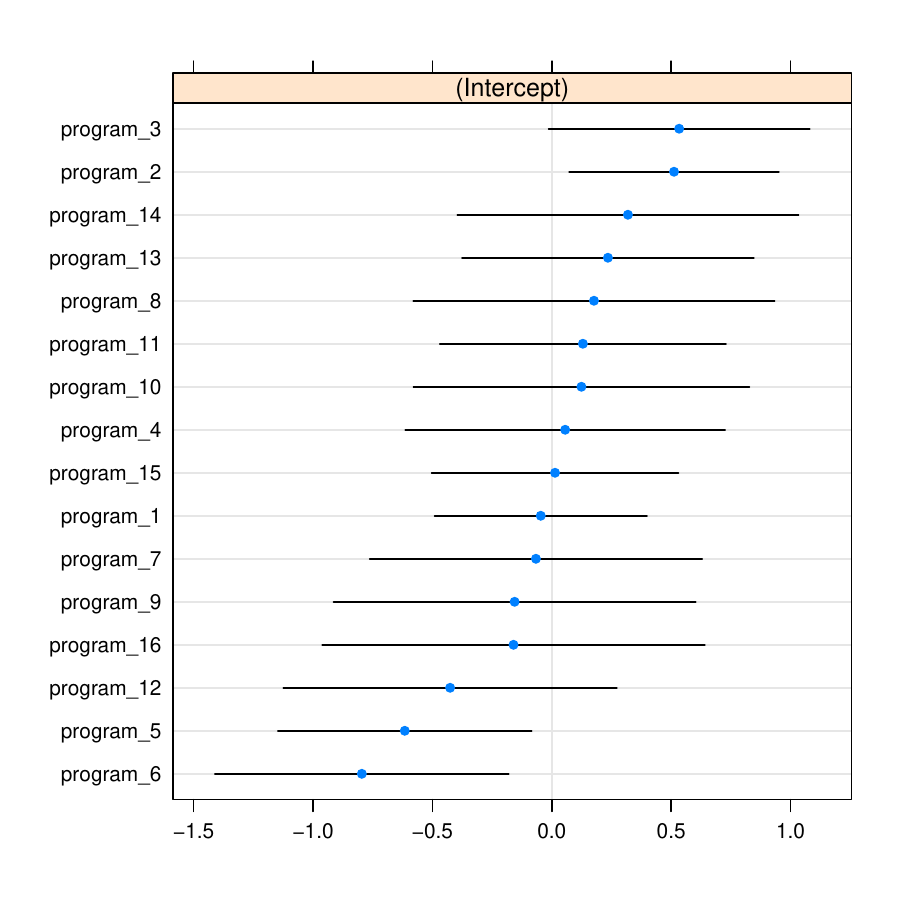}}}
    \caption{Estimated coefficients of the model with binary output (Mod. \ref{eq:model_binary}).} \label{fig:Binary_output}
\end{figure}

The model successfully classifies 508 out of 669 students using LOO-CV, achieving an accuracy of 76\%. Figure \ref{fig:Binary_output}a presents the Odds Ratio values, indicating that both high school grade and the specific high school attended are statistically significant variables.
Although the differences in other variables, it is important to note that the variable of main interest, i.e., commuting time, is not statistically significant.

In light of the exploratory analysis involving descriptive statistics, comparisons of sample distributions, and model evaluations, it can be confidently concluded that the sample is appropriate to describe students living in the metropolitan area of Milan.

\end{document}